\documentclass{aa}  
\usepackage{graphicx}
\usepackage{txfonts}
\begin{document}
   \title{The radio structure of radio--quiet quasars}

   \subtitle{}

   \author{C. Leipski
          \inst{1}
          \and
          H. Falcke
	  \inst{2,3}
          \and
          N. Bennert
          \inst{1,4}
          \and
          S. H\"uttemeister
          \inst{1}  }

   \offprints{C. Leipski}

   \institute{AIRUB, University of Bochum, Universit\"atsstrasse 150,
              44780 Bochum, Germany\\
              \email{leipski@astro.rub.de; nbennert@astro.rub.de;
                     huette@astro.rub.de}
             \and
             Radio Observatory Westerbork, ASTRON, P.O. Box 2, 7990 AA
              Dwingeloo, The Netherlands\\
             \email{falcke@astron.nl}
             \and
	     Department of Astrophysics, Radboud University, 6525 ED
              Nijmegen, The Netherlands
             \and
             Institute of Geophysics and Planetary Physics, University
              of California, Riverside, CA 95521, USA\\
             \email{nicola.bennert@ucr.edu}}

   \date{Received October 6, 2005; accepted April 26, 2006}

% \abstract{}{}{}{}{} 
% 5 {} token are mandatory
 
  \abstract
   {}
   {We investigate the radio emitting structures of radio--quiet
  active galactic nuclei with an emphasis on radio--quiet quasars
  to study their connection to Seyfert galaxies.}
   {We present and analyse high--sensitivity VLA radio continuum images of 14
  radio--quiet quasars and six Seyfert galaxies.}
   {
   Many of the low redshift radio--quiet quasars
   show radio structures that can be interpreted as jet--like outflows.
   However, the detection rate
  of extended radio structures on arcsecond scales among our
  sample decreases with increasing redshift and luminosity, most
  likely due to a lack of resolution.
  The morphologies of the
  detected radio emission indicate strong interactions 
  of the jets with the surrounding medium. We also compare the radio
  data of seven 
  quasars with corresponding HST images of the [\ion{O}{iii}] emitting
  narrow--line region (NLR). We find that the scenario of interaction
  between the radio jet and the NLR gas is confirmed
  in two sources by structures in
  the NLR gas distribution as previously known for Seyfert galaxies. The
  extended radio structures of radio--quiet quasars at
  sub--arcsecond
  resolution are by no means different from that 
  of Seyferts. Among the luminosities studied here, the morphological
  features found are similar in both types of objects while the overall
  size of the radio structures increases with luminosity. This
  supports the picture where radio--quiet quasars are the scaled--up
  versions of 
  Seyfert galaxies. In addition to known luminosity relations we find
  a correlation of the NLR size and the radio size shared by quasars and
  Seyferts.
   } 
   {}

   \keywords{Galaxies: active -- Galaxies: jets -- Galaxies: Seyfert
               }

   \maketitle
%
%________________________________________________________________

\section{Introduction}
         The population of active galactic nuclei (AGN) shows a well
         known wide spread in radio luminosity (Miller et al. \cite{miller90};
         Kellermann et al. \cite{kellermann89},\cite{kellermann94};
         Miller et al. \cite{miller93}; Kukula et
         al. \cite{kukula98}; Xu et al. \cite{xu99}).
         Introducing the R parameter as the
         ratio of radio flux at 5 GHz and optical flux at 4400\,\AA~
	 (Kellermann et al. \cite{kellermann89}; Falcke et
         al. \cite{falcke96b}), 
         the radio--loud objects
         are separated  from the radio--quiet objects: Usually,
         the radio--loud quasars have R\,$>$\,100 and 5 GHz
         luminosities greater than $10^{26}\,{\rm W\,Hz^{-1}}$. The
         radio--quiet ones typically have R values between 0.1 and 10 and
         luminosities less than $10^{24}\,{\rm W\,Hz^{-1}}$ at 5 GHz.
         Thus, the separation between
         the classes is nearly three orders of magnitude in radio
         luminosity (Kellermann et
         al. \cite{kellermann89},\cite{kellermann94}). Radio--loud
         quasars (RLQs) and radio--galaxies are usually dominated by prominent 
         double--lobe structures extending to tenth or hundreds of
         kpc. These extended structures in radio--loud
         objects are also dominating in luminosity by outshining the
         emission from the active central region by factors of 10 to
         100, resulting in a very low core--to--lobe flux
         ratio. On the other hand, radio--quiet quasars (RQQs)
         only show small scale structures of typically a few
         kpc with most of the flux concentrated in the nuclear region
	 (e.g. Miller et al. \cite{miller90}; Kellermann et al.
         \cite{kellermann94}).
	 Nevertheless, some RQQs are known to
         show extended structures reminiscent of radio--loud
         objects, but on very low flux--density scales 
         (Kellermann et al. \cite{kellermann94}; Blundell \& Rawlings
         \cite{blundell01}). However, systematic high--dynamic
         range radio imaging studies of RQQs are still missing.\\\indent
         At lower (radio) luminosities, Seyfert galaxies also show 
         compact structures, often dominated by a
         central unresolved component, sometimes accompanied by
         jet--like features (Ulvestad et al. \cite{ulvestad81}; Kukula
         et al. \cite{kukula95}; Schmitt et al. \cite{schmitt01}; Ho \&
         Ulvestad \cite{ho01}). Comparing the snapshot data of Seyfert
         galaxies and RQQs, there is no significant difference in the radio
         structure in both types of radio--quiet AGN. \\\indent
         However, in high sensitivity data, the diversity in radio
         structure of Seyfert galaxies is obvious (Ulvestad \& Wilson
         \cite{ulvestad89}; Capetti et al. \cite{capetti96}; Falcke et al.
         \cite{falcke98}). The morphologies can comprise linearly arranged
         multiple components (interpreted as emission knots in a
         jet), one-- or two--sided diffuse jets or
         complicated structures that are, nevertheless, often well
         collimated. These radio structures coincide in shape and/or direction
         with the line--emitting gas
         of the narrow--line region (NLR), indicating
         that the radio ejecta are intimately connected to
         the NLR gas (Falcke et al. \cite{falcke98}).\\\indent
	 In the ongoing debate of the evolution of active galaxies
         with luminosity, these findings of structured NLRs in Seyfert
         galaxies lead Bennert et al. (\cite{bennert02}) to investigate
         the NLR structure of RQQs with high--resolution. 
         They find resolved and structured quasar
         NLRs that closely resemble that of Seyfert galaxies. Moreover
         they discovered a close correlation between the size and the
         luminosity of the NLR, connecting Seyfert galaxies and RQQs
         over a few orders of magnitude in luminosity.
         These results strongly support the idea that, amongst
         radio--quiet AGN, RQQs are the high--power
         siblings of Seyfert galaxies. Since the close connection of
         Seyferts and RQQs is thus supported in the optical/NLR, we here 
	 explore the question if this can in fact be extended to the
         radio regime. We investigate if the interaction of radio
         ejecta and NLR gas, a typical phenomenon of active galaxies,
         persists with increasing luminosity. Therefore, we combine
         high--resolution, high--sensitivity radio and optical data
         for a sample of RQQs. We present radio continuum maps
         for 14 RQQs and some Seyfert galaxies to analyse the
         similarities and differences in both types of objects.\\\indent
         In $\S$2 and $\S$3, observations, data reduction, and data
         analysis are summarised. In $\S$4, we present the results on
         an object--by--object basis. In $\S$5, we discuss the
         results and their implications for the role of radio
         jets. We close with the concluding remarks in $\S$6.
\begin{table*}
%\resizebox{\hsize}{!}{
%\begin{minipage}{180mm}
%\begin{center}
\caption[]{\sc VLA Observation Log
         \label{sample}}
\begin{center}
%\resizebox{\hsize}{!}{
\begin{tabular}{cllllccc}
\hline\hline \\[-2.5ex]
 & & \multicolumn{2}{c}{optical position} & & & Frequency & on--source \\[0.5ex]\cline{3-4}
& \multicolumn{1}{c}{Object} & \multicolumn{1}{c}{\raisebox{0.2ex}[-0.2ex]{$\alpha_{2000}$}} & \multicolumn{1}{c}{\raisebox{0.2ex}[-0.2ex]{$\delta_{2000}$}} & \hspace*{+1mm}z$_{\rm hel}$ & R & (GHz) & time \\ 
& \multicolumn{1}{c}{(1)} & \multicolumn{1}{c}{(2)} & \multicolumn{1}{c}{(3)} & \hspace*{+1mm}(4) & (5) & (6) & (7)\\*[0.1cm]\hline
& PG0026+129    & 00 29 13.73  & +13 16 04.3    & 0.142    & 1.10 & 8.4 & 2:02 \\[0.3ex]
& PG0052+251    & 00 54 52.16  & +25 25 38.7    & 0.155    & 0.24 & 4.8 & 3:57 \\[0.3ex]
& PG0157+001    & 01 59 50.21  & +00 23 40.6    & 0.163    & 2.10 & 8.4 & 2:22 \\[0.3ex]
\multicolumn{1}{c}{\rotatebox{0}{Quasars}} & PG0953+414    & 09 56 52.50  & +41 15 40.6    & 0.2341   & 0.44 & 4.8 & 3:45 \\[0.3ex]
& PG1012+008    & 10 14 54.88  & +00 33 36.8    & 0.1874   & 0.50 & 8.4 & 2:46 \\[0.3ex]
& PG1049$-$005  & 10 51 51.50  & $-$00 51 16.6  & 0.3599   & 0.25 & 4.8 & 3:34 \\[0.3ex]
& PG1307+085    & 13 09 47.03  & +08 19 49.3    & 0.155    & 0.10 & 4.8 & 3:07 \\[0.3ex]\hline
& PG0003+199    & 00 06 19.193 & +20 12 10.51   & 0.0258   & 0.27 & 4.8 & 2:59 \\[0.3ex]
& PG1119+120    & 11 21 47.102 & +11 44 18.28   & 0.0502   & 0.15 & 4.8 & 2:33 \\[0.3ex]
& PG1149$-$110  & 11 52 03.505 & $-$11 22 23.87 & 0.049    & 0.88 & 4.8 & 3:05 \\[0.3ex]
\multicolumn{1}{c}{\rotatebox{0}{Quasars}} & PG1351+640    & 13 53 15.847 & +63 45 45.72   & 0.0882   & 4.30 & 4.8 & 3:54 \\[0.3ex]
& PG1534+580    & 15 35 52.358 & +57 54 09.17   & 0.0296   & 0.70 & 4.8 & 3:47 \\[0.3ex]
& PG1612+261    & 16 14 13.214 & +26 04 16.40   & 0.131    & 2.80 & 4.8 & 3:17 \\[0.3ex]
& PG2130+099    & 21 32 27.814 & +10 08 19.49   & 0.063    & 0.32 & 4.8 & 2:12 \\[0.3ex]\hline
& Mrk 612       & 03 30 40.9   & $-$03 08 16    & 0.0203   & & 8.4 & 1:30 \\[0.3ex]
& ESO 428$-$G14 & 07 16 31.2   & $-$29 19 28    & 0.0054   & & 8.4 & 1:42 + 1:05 \\[0.3ex]
& NGC 2639      & 08 43 38.1   & +50 12 20      & 0.0111   & & 8.4 & 1:06 \\[0.3ex]
\multicolumn{1}{c}{\raisebox{1.8ex}[-1.8ex]{Seyferts}} & NGC 2992      & 09 45 42.0   & $-$14 19 35    & 0.0077   & & 8.4 & 1:06 \\[0.3ex]
& Mrk 266       & 13 38 17.5   & +48 16 37      & 0.0279   & & 8.4 & 1:39 \\[0.3ex]
& NGC 5643      & 14 32 40.8   & $-$44 10 29    & 0.004    & & 8.4 & 1:47 + 2:19 \\*[0.1cm]\hline
%NGC 6500      & 17 55 59.8   & +18 20 18      & 0.0100 & 8.4 & 0 43 \\*[0.1cm]\hline
\end{tabular}
\end{center}
%\end{minipage}
\begin{list}{}{}
\item[\emph{Note:}] The first seven RQQs are also included in the HST sample of
  Bennert et al. (\cite{bennert02}).
  %\item[(2)--(3)] The first seven RQQ are also included in the HST data of
%  Bennert et al. (\cite{bennert02}).
\item[(2)--(3)] The optical position was taken from Miller et al. \cite{miller93} and calculated for J2000 (in $^{\rm h}\,^{\rm min}\,^{\rm sec}$ and $^{\circ}\,^{\prime}\,^{\prime\prime}$).
\item[(4)] heliocentric redshift as provided by NED
  \item[(5)] R parameter taken from Kellermann et al. (\cite{kellermann94})
\item[(7)] on--source integration time in hrs:min
\end{list}
%\end{center}
\end{table*}
\section{Observations and data reduction}
\begin{table*}
%\resizebox{\hsize}{!}{
%\begin{minipage}{180mm}
%\begin{center}
\caption[]{\sc Source properties
         \label{sample_results}}
\begin{center}
%\resizebox{\hsize}{!}{
\begin{tabular}{lllcccccccc}
\hline\hline \\[-2.5ex]
 & \multicolumn{2}{c}{peak position} & peak flux & total flux & size &
 $3\,\sigma_{\rm rms}$ & total flux$_{\,\rm B}$ &
 \multicolumn{2}{c}{Luminosity} & scale \\[0.5ex]\cline{2-3}\cline{9-10}
Object & \multicolumn{1}{c}{\raisebox{0.2ex}[-0.2ex]{$\alpha_{2000}$}}
 & \multicolumn{1}{c}{\raisebox{0.2ex}[-0.2ex]{$\delta_{2000}$}} &
 \multicolumn{1}{c}{mJy} & \multicolumn{1}{c}{mJy} & in
 $^{\prime\prime}$ & $10^{-5}\,{\rm Jy}$ & mJy &
 \multicolumn{2}{c}{$10^{21}\,\frac{\rm W}{\rm Hz}$} & ${\rm pc}/^{\prime\prime}$\\ 
(1) & \multicolumn{1}{c}{(2)} & \multicolumn{1}{c}{(3)} & (4) & (5) &
 (6) & (7) & (8)  & (9) & (10) & (11) \\*[0.1cm]\hline
PG0026+129    & 00 29 13.700 & +13 16 04.00   & 0.17   & 0.28   & 0.6     & 4.5 & 2.37 & 14.52  & 129.95 & 2442.6 \\[0.3ex]
PG0052+251    & 00 54 52.116 & +25 25 39.03   & 0.58   & 0.61   & $<0.09$ & 3   & -    & 38.45  &        & 2631.8 \\[0.3ex]
PG0157+001    & 01 59 50.252 & +00 23 40.89   & 2.98   & 4.23   & 2.6     & 3   & 5.55 & 298.74 & 391.96 & 2746.7 \\[0.3ex]
PG0953+414    & 09 56 52.400 & +41 15 22.16   & 0.13   & 0.18   & 1.3     & 3   & 0.18 & 29.16  & 29.16  & 3684.0 \\[0.3ex]
PG1012+008    & 10 14 54.900 & +00 33 37.46   & 0.10   & 0.25   & 1.0     & 3   & 0.65 & 24.72  & 64.28  & 3106.6 \\[0.3ex]
PG1049$-$005  & 10 51 51.446 & $-$00 51 17.70 & 0.22   & 0.25   & $<0.27$ & 3   & -    & 109.57 &        & 4987.8 \\[0.3ex]
PG1307+085    & 13 09 47.000 & +08 19 48.16   & 0.13   & 0.15   & 1.1     & 3   & -    & 9.74   &        & 2661.6 \\[0.3ex]\hline
PG0003+199    & 00 06 19.537 & +20 12 10.61   & 3.49   & 3.58   & $<0.08$ & 5   & -    & 4.69   &        & 481.4  \\[0.3ex]
PG1119+120    & 11 21 47.122 & +11 44 18.93   & 0.08   & 0.39   & 5.6     & 3   & 1.04 & 2.38   & 6.34   & 985.7  \\[0.3ex]
PG1149$-$110  & 11 52 03.549 & $-$11 22 24.10 & 1.08   & 1.94   & 2.4     & 5.5 & -    & 11.28  &        & 964.8  \\[0.3ex]
PG1351+640    & 13 53 15.828 & +63 45 45.52   & 7.31   & 8.48   & 3.5     & 3   & -    & 161.70 &        & 1626.8 \\[0.3ex]
PG1534+580    & 15 35 52.400 & +57 54 09.53   & 2.00   & 2.21   & $<0.22$ & 5   & -    & 4.36   &        & 584.8  \\[0.3ex]
PG1612+261    & 16 14 13.221 & +26 04 16.27   & 1.19   & 5.27   & 2.6     & 5.5 & 6.12 & 234.28 & 272.07 & 2299.8 \\[0.3ex]
PG2130+099    & 21 32 27.816 & +10 08 19.25   & 0.68   & 2.41   & 2.9     & 6   & -    & 21.61  &        & 1172.3 \\[0.3ex]\hline
Mrk 612       & 03 30 40.906 & $-$03 08 14.56 & 0.19   & 0.68   & 1.9     & 3   & -    & 0.59   &        & 393.8  \\[0.3ex]
ESO 428$-$G14 & 07 16 31.165 & $-$29 19 28.37 & 3.84   & 18.95  & 4.8     & 4   & -    & 1.50   &        & 122.6  \\[0.3ex]
NGC 2639      & 08 43 38.078 & +50 12 19.98   & 132.16 & 152.97 & 2.0     & 7   & -    & 44.80  &        & 233.4  \\[0.3ex]
NGC 2992      & 09 45 41.946 & $-$14 19 34.67 & 4.81   & 12.00  & 7.6     & 4.5 & -    & 2.03   &        & 178.3  \\[0.3ex]
Mrk 266       & 13 38 17.786 & +48 16 41.16   & 3.47   & 15.00  & 12.4    & 5   & -    & 27.06  &        & 560.5  \\[0.3ex]
NGC 5643      & 14 32 40.703 & $-$44 10 27.41 & 5.68   & 11.70  & 41.0    & 4   & -    & 0.55   &        & 94.7   \\*[0.1cm]\hline
%NGC 6500      & 17 55 59.8   & +18 20 18      &  &  &  & \\*[0.1cm]\hline
\end{tabular}
\end{center}
%\end{minipage}
\begin{list}{}{}
\item[\emph{Note:} (2)--(3)] in $^{\rm h}\,^{\rm min}\,^{\rm sec}$ and $^{\circ}\,^{\prime}\,^{\prime\prime}$
\item[(6)] major extension (diameter) of source emission, a "$<$"
  means that only an upper limit can be estimated
\item[(8)] B--Array flux
\item[(9)--(10)] luminosity from the A and B--Array fluxes, respectively
\item[(4)--(10)] flux, size, $3\,\sigma_{\rm rms}$, and luminosity at
  the observed frequency shown in Tab.\,\ref{sample}
\end{list}
%\end{center}
\end{table*}
         This paper makes use of four different samples that we 
         now present briefly: First, the quasar sample of Bennert et
         al. (\cite{bennert02}). The
         objects are the seven most--luminous (in [\ion{O}{iii}])
         quasars with $z<0.5$ in the Palomar Green (PG) quasar
         sample (Schmidt \& Green \cite{schmidt83}). Only for
         $z<0.5$ the PG quasar sample is complete and not affected by
         a selection bias (e.g. Wampler \& Ponz \cite{wampler85}).
         Only for this sub--sample HST, emission--line images are
         available.
	 
	 The second sample consists of seven PG quasars with
         redshift $z$\,$<$\,0.2 and [\ion{O}{iii}] emission line
         fluxes $>$ 5 $\cdot$ 10$^{-14}$ erg s$^{-1}$ cm$^{-2}$.
	 This corresponds to [\ion{O}{iii}] luminosities exceeding those
         for typical Seyfert
         galaxies, thereby filling the gap between quasars and Seyferts in
         the NLR size-luminosity correlation of Bennert et
         al. (\cite{bennert02}).
	 These objects were selected as ``transition objects''
	 in terms of their [\ion{O}{iii}] luminosities and (expected)
         NLR sizes and are thus of special interest.
	 
	 Third, a sample of six Seyfert galaxies that were selected on
         the basis of their extended morphological appearance in
         radio snapshot surveys. This sample is presented in the
         appendix.
	 
         For all these objects, the basic parameters of the
         observations are given in Tab.\,\ref{sample}.
         In $\S$5, we will make use of the Seyfert sample of Kinney et
         al. (\cite{kinney00}) and Schmitt et al. (\cite{schmitt01},
         \cite{schmitt03}). These objects were selected on the basis
         of a mostly isotropic property, their 60\,$\mu$m fluxes and
         their warm infrared colours.\\\indent
	 The data for the first three samples were newly obtained using the
         Very Large Array (VLA\footnote{The National Radio Astronomy
         Observatory is a facility of the National Science Foundation
         operated under cooperative agreement by Associated
         Universities, Inc.}; 
         Thompson et al. \cite{thompson80}) in the A, BnA, and B configuration
         at 4.8 GHz or 8.4 GHz with two intermediate
         frequencies (IFs) of 50 MHz bandwidth. The details of the
         observations are given in Table 1.
	 The data reduction
         was performed in the typical manner using the {\sc Aips}
         software. All bright sources were subject to several
         iterations of self--calibration.
         Natural-- and uniform--weighted (with a robust parameter of
         -4) maps were
         produced with various cleaning depths, taperings and {\sc Clean}
         gain factors.\\\indent
         In $\S$4 we present and discuss the radio maps of the newly
         observed objects 
         on an object--by--object basis. The maps of the quasars show
         compact cores, often accompanied by weak extended emission or
         additional unresolved components. All maps in this paper
         are shown in contours with spacings of
	 ${\sqrt{2}}^{\,\rm n}$, starting at $3\,\sigma_{\rm rms}$ as
         given in Table 2. 
         \section{Data analysis}
         Peak positions of the radio sources were determined in the
         high resolution uniform--weighted maps by fitting a
         two--dimensional {\sc Gauss}--function to the central peak
         using the {\sc Aips}--task {\sc Imfit}. For diffuse and
         extended sources for which a multi--component {\sc
         Gaussian}--fit is not suited to represent the source structure
         the total flux was derived by summing the flux inside the
         $3\sigma_{\rm rms}$ contour.
         The noise in the map was determined over large areas near the
         source using the task 
         {\sc Tvstat}. \\
         The diameter of the radio sources was determined by measuring the
         widest extent of the source in the $3\,\sigma_{\rm rms}$ contour.
         This diameter was corrected for the
         elongation of the beam. However, this correction has only a minor
         impact on the overall source size and becomes less important with
         increasing total extent of the source. For the compact sources, the
         deconvolution of the beam shape and the apparent source shape
         was performed by the task {\sc Imfit}. For all cases in which a
         source appeared compact, it was only marginally resolved after
         deconvolution. We take this marginal extent only as
         an upper limit on the source size.\\
         From the fluxes and diameters, luminosities
         and linear distances were calculated using a homogeneous flat
         world model which includes Einstein's cosmological constant
         $\Lambda$ in agreement with the recent
         results of the Wilkinson Microwave Anisotropy Probe (Bennet
         et al. \cite{bennet03}: $H_0=71\,{\rm km\,}{\rm s}^{-1}$, 
         $\Omega_{{\rm matter}}$ = 0.27 and $\Omega_{\Lambda}$ =
         0.73). Results from the data analysis are given in Tab.\,\ref{sample_results}.
         \begin{figure*}
         \centering
         \includegraphics[width=6cm,angle=-90]{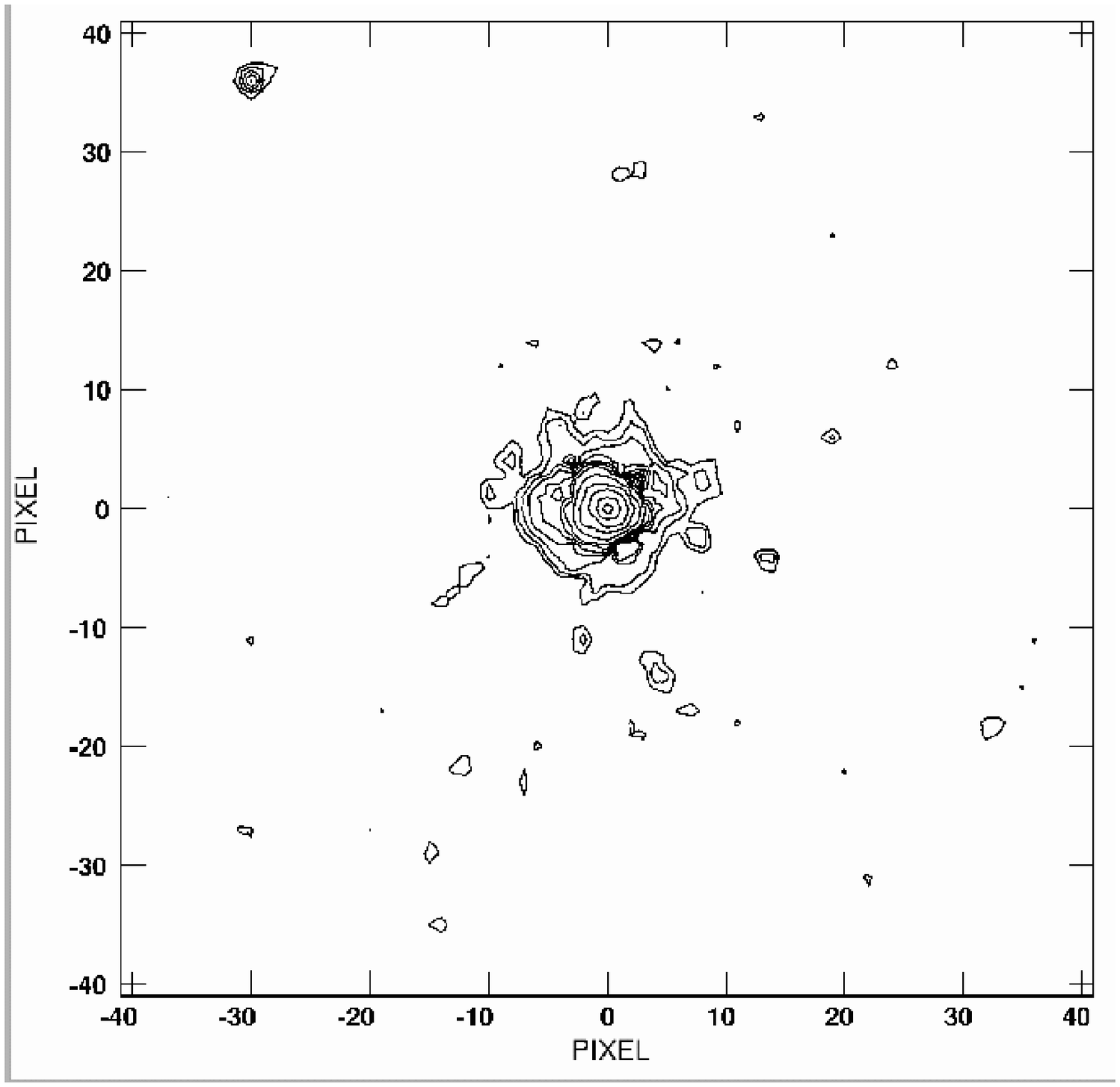}
         \includegraphics[width=6cm,angle=-90]{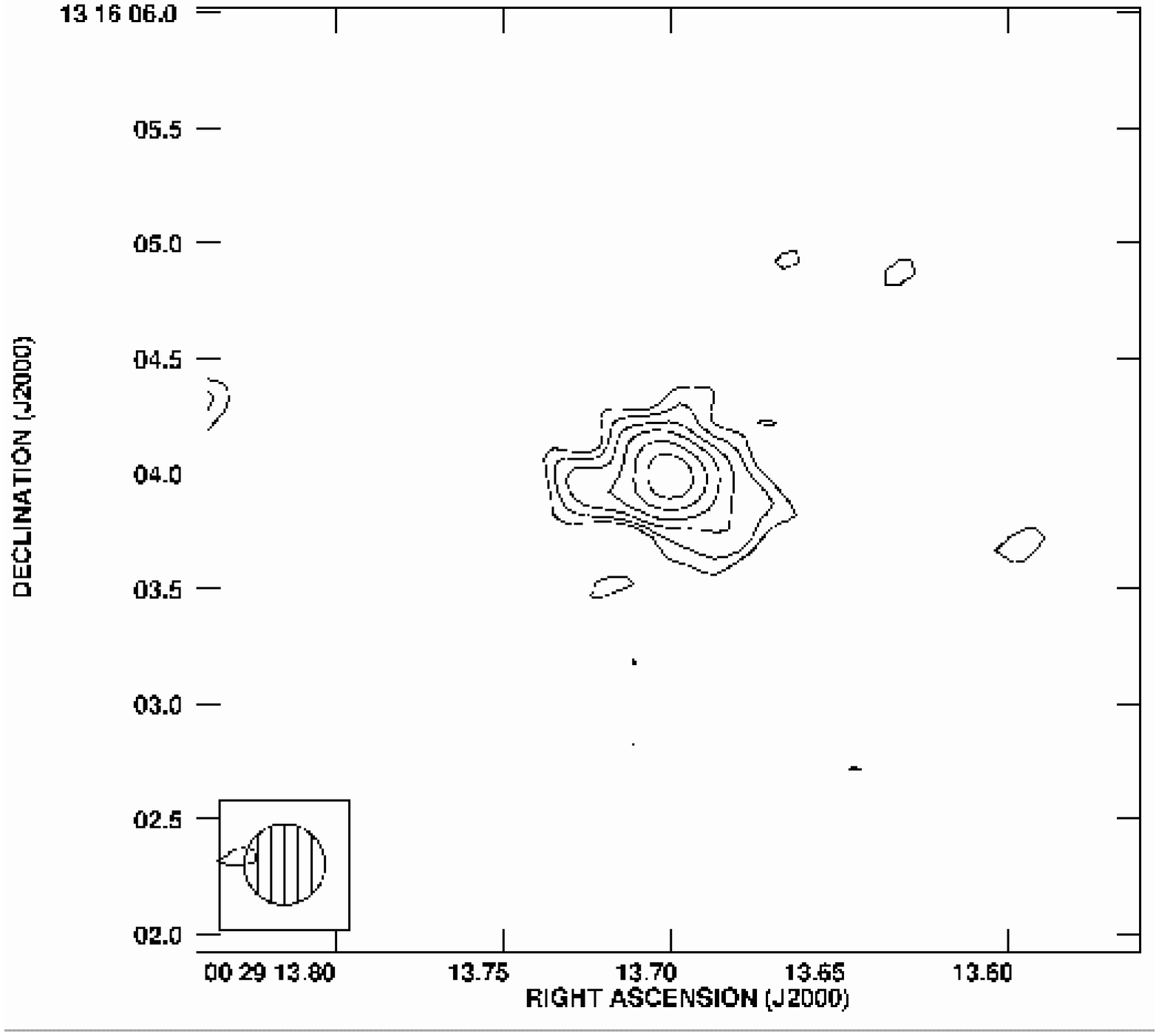}\\\hspace*{-0.5cm}
         \includegraphics[width=6cm,angle=-90]{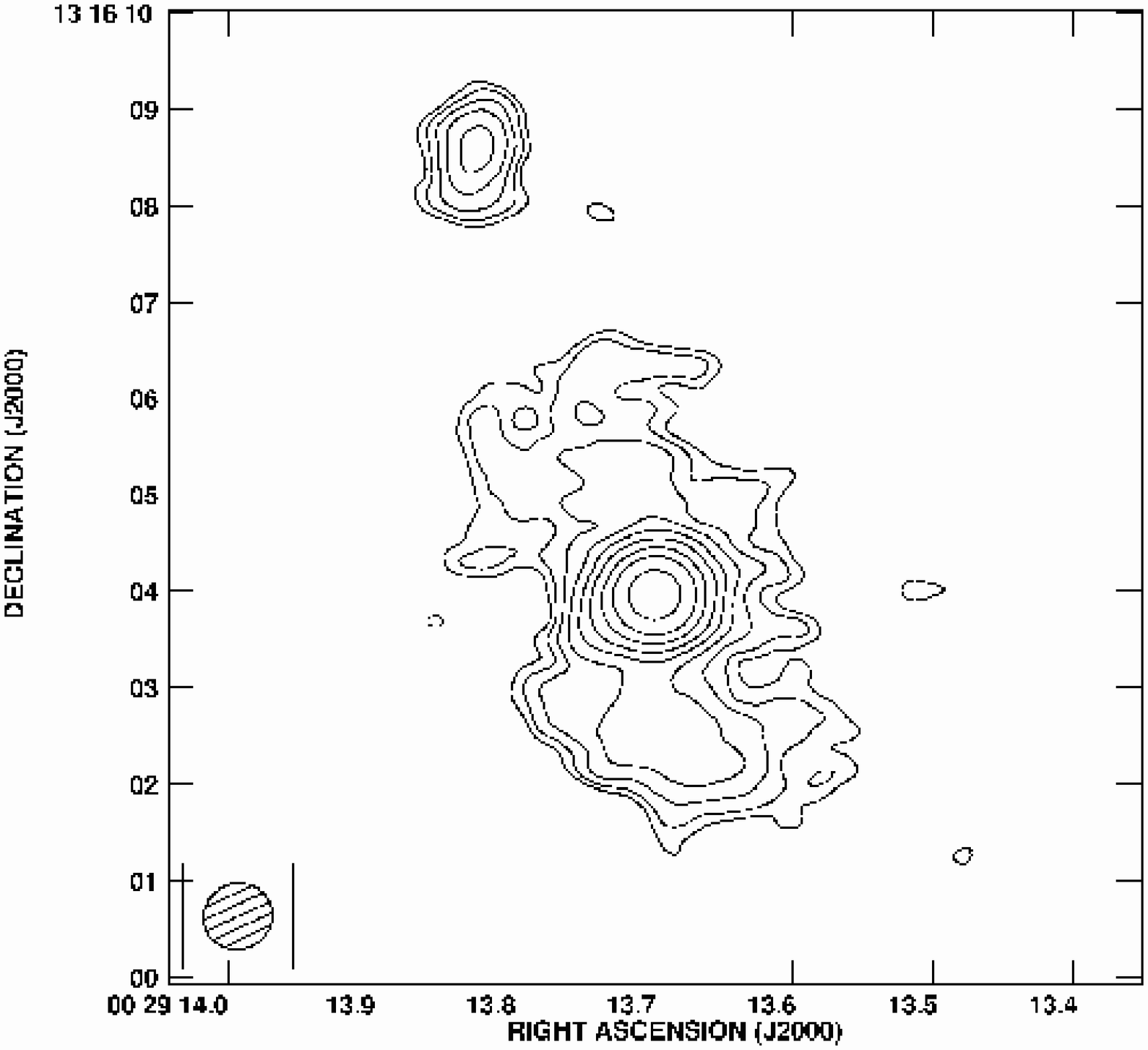}
         \includegraphics[width=6cm,angle=-90]{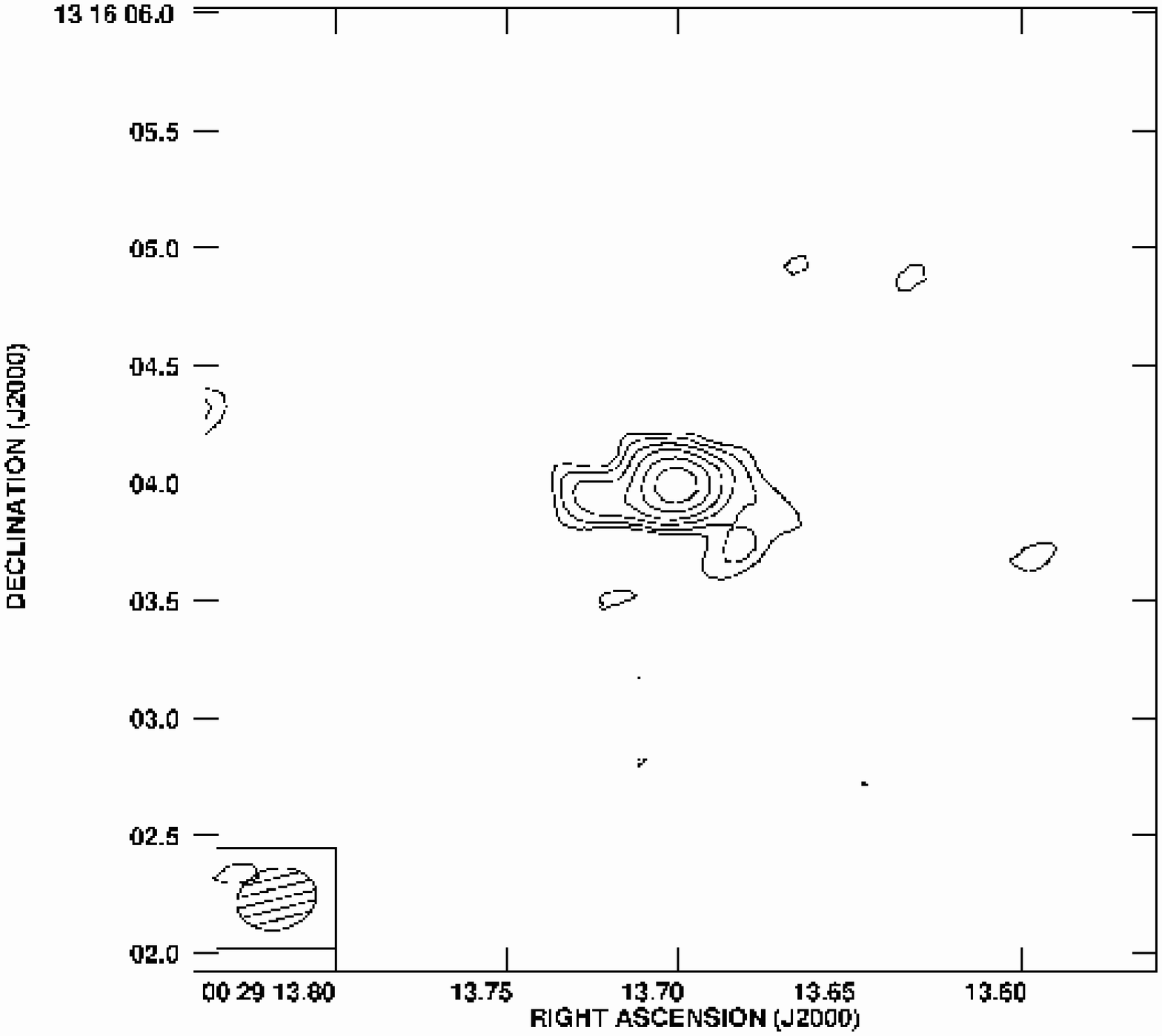}
         \caption[]{\label{pg0026}HST and VLA images of
         PG0026+129. Images are
         $4^{\prime\prime}\times4^{\prime\prime}$ wide, if not stated
         otherwise. {\sl Upper left}: HST linear ramp filter (LRF) image in the
         [\ion{O}{III}] line. Pixel offsets
         from centre correspond to
         0.05$\,^{\prime\prime}/{\rm px}$. {\sl Upper right}:
         Natural--weighted VLA
         A--Array map at 8.4 GHz with a $0.^{\hspace*{-0.1cm}\prime\prime}35$
         beam. {\sl Lower left}: Natural--weighted VLA B--Array map at
         8.4 GHz with 
         a 500 k$\lambda$ gaussian taper (beamsize
         $0.^{\hspace*{-0.1cm}\prime\prime}72\times0.^{\hspace*{-0.1cm}\prime\prime}68$).    
         The size of the image is $10^{\prime\prime}$.
         {\sl Lower right}: Natural--weighted VLA
         A--Array map at 8.4 Ghz (beamsize
         $0.^{\hspace*{-0.1cm}\prime\prime}34\times0.^{\hspace*{-0.1cm}\prime\prime}26$).
         }   
\end{figure*}
         \section{Results}
	 \begin{figure*}
         \centering
         \includegraphics[width=6cm,angle=-90]{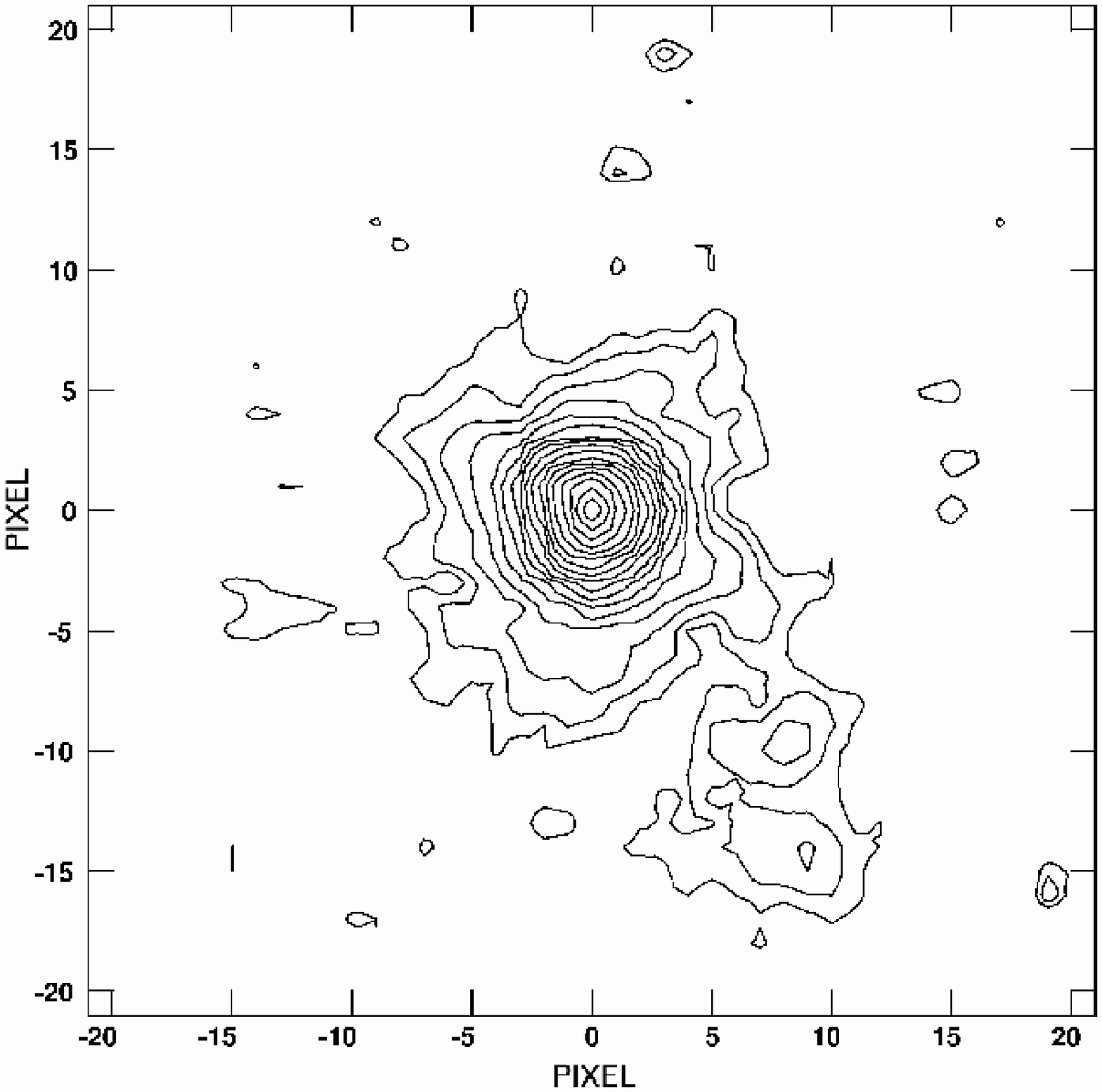}
         \includegraphics[width=6cm,angle=-90]{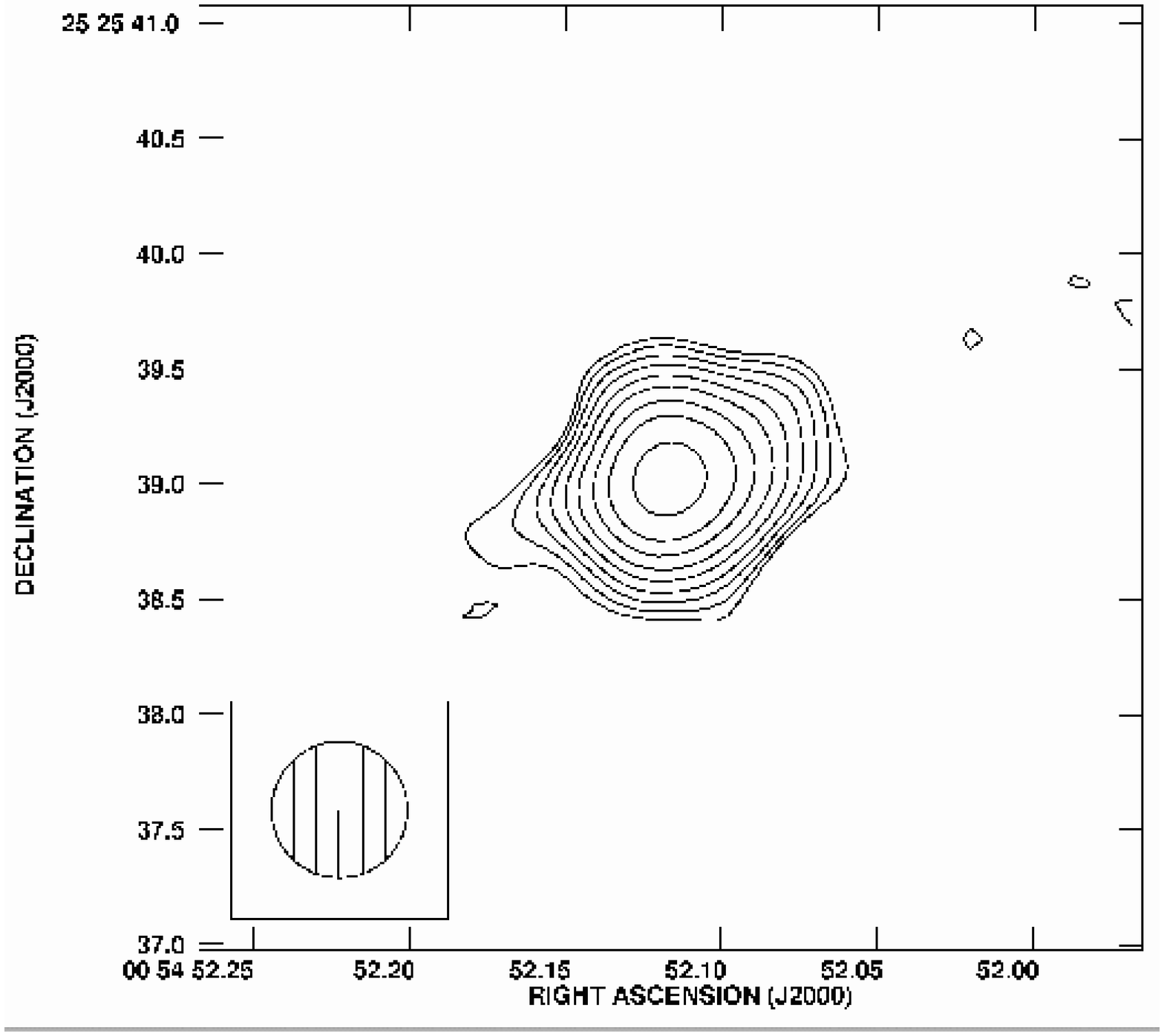}
         \caption[]{\label{pg0052}HST and VLA images of
         PG0052+251. Both images are
         $4^{\prime\prime}\times4^{\prime\prime}$ wide.
         {\sl Left}: HST LRF image in the
         [\ion{O}{iii}] line. Pixel offsets
         from centre correspond to
         0.1$\,^{\prime\prime}/{\rm px}$. {\sl Right}:
         Natural--weighted VLA
         A--Array map at 4.8 GHz with a 
         beam of $0.^{\hspace*{-0.1cm}\prime\prime}59$. }
         \end{figure*}
\begin{figure*}
         \centering
         \includegraphics[width=6cm,angle=-90]{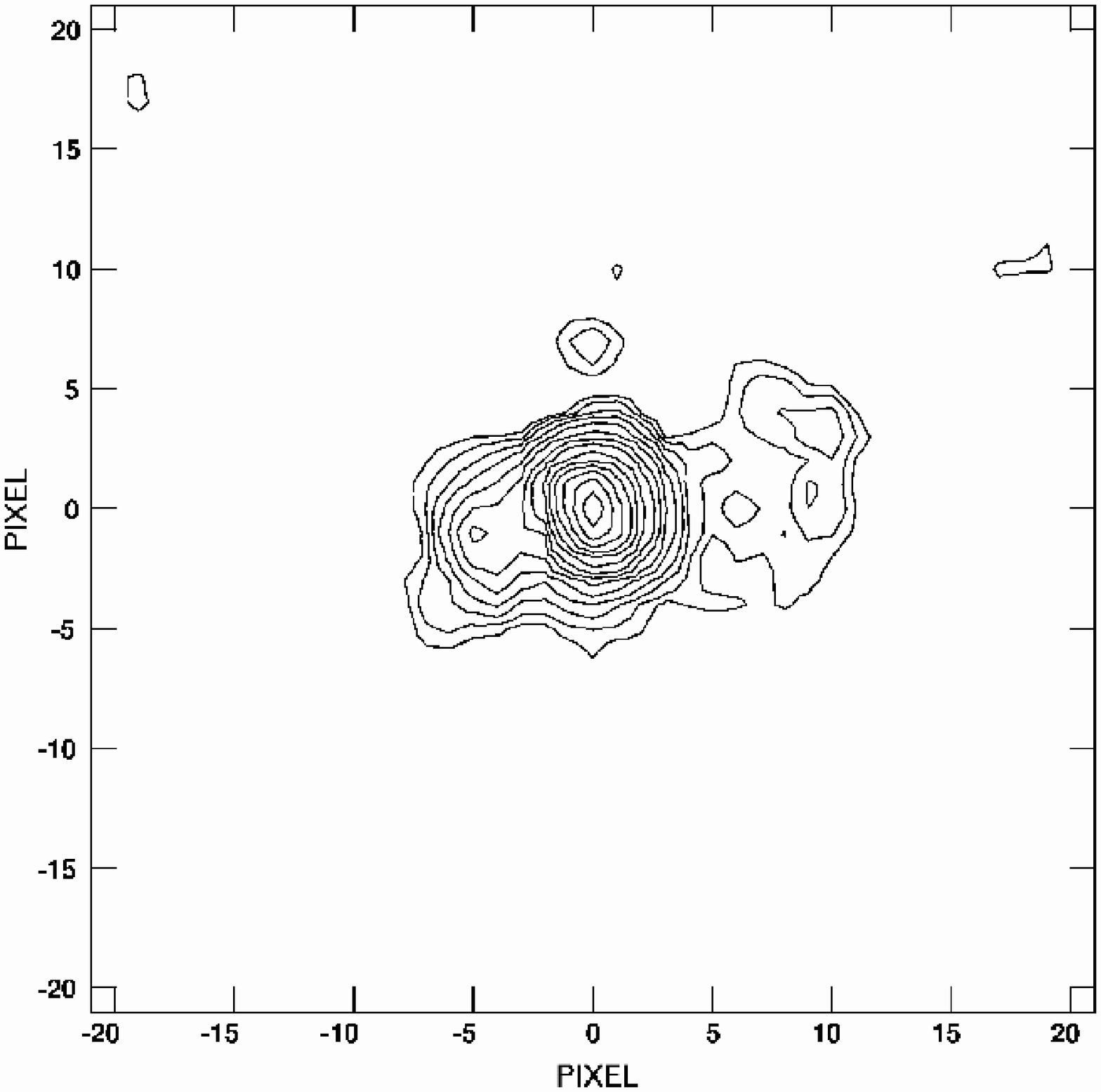}
         \includegraphics[width=6cm,angle=-90]{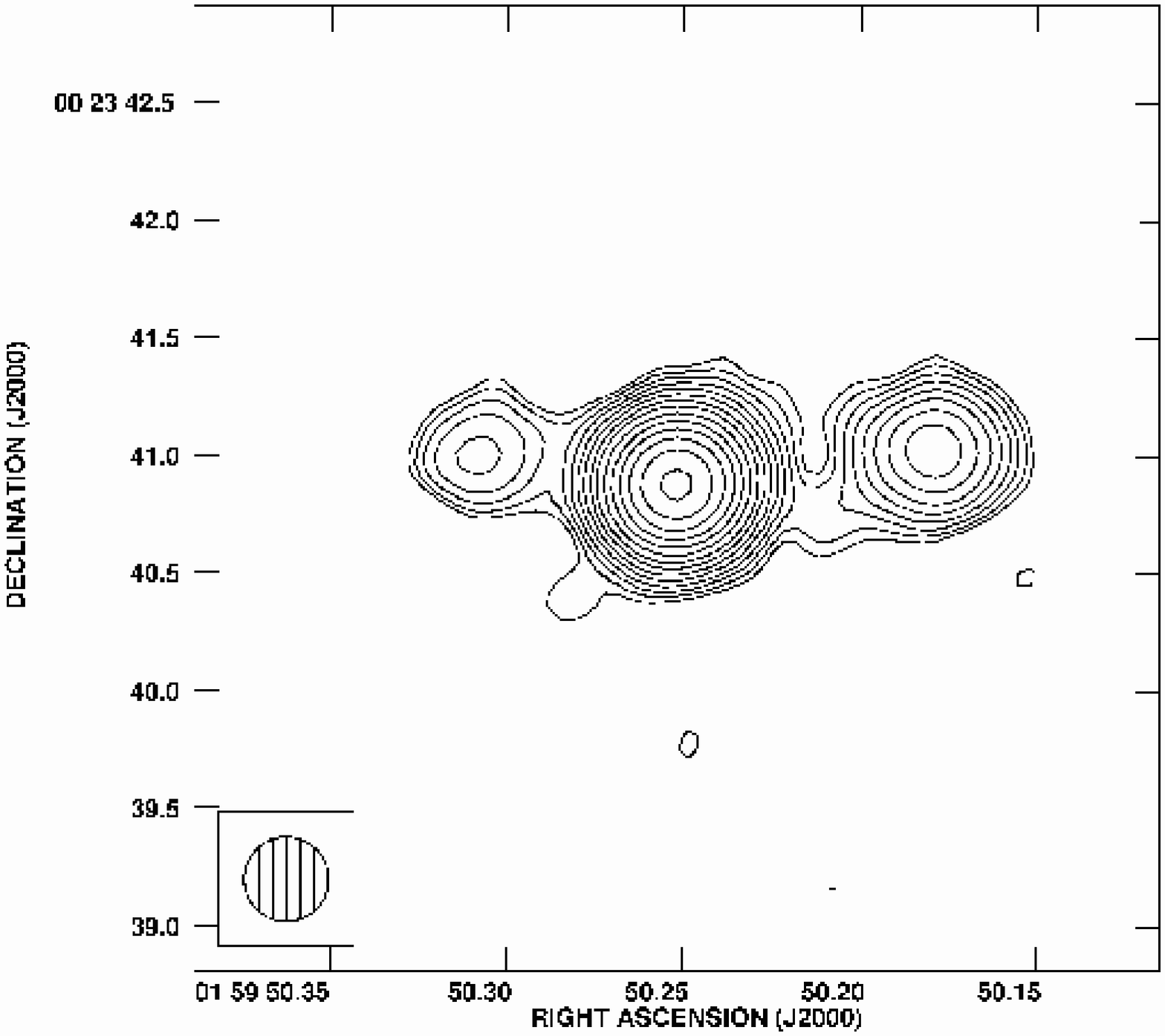}
         \caption[]{\label{pg0157}Same as in Fig.\,2 for
         PG0157+001. The VLA map is at 8.4 GHz with a $0.^{\hspace*{-0.1cm}\prime\prime}36$ beam. }
         \end{figure*}    

         \subsection{Radio--quiet quasars}
         The radio--quiet quasars presented here consist of
         two groups, each with seven objects, and all taken from the
         PG quasar sample (Schmidt \& Green \cite{schmidt83}).
	 The seven objects in the first part of Tabs.\,\ref{sample}
	 and \ref{sample_results} have been imaged with
         HST in the redshifted 
         [\ion{O}{iii}] line by Bennert et al. (\cite{bennert02}).\\\indent
         Their emission--line images and our new radio maps are shown
         below.
         Table 2 lists the central coordinates, integrated and peak fluxes, 
         as well as sizes and noise levels in the maps. Derived
         luminosities and linear source scales are also given.
	 \subsubsection{Seven RQQs with HST NLR images}
	 
	 \indent{\bf PG0026+129}. This source shows no remarkable
         extent in neither 
         the HST nor the high--resolution VLA image
         (Fig. 1). Kellermann et al. (1989) 
         found that only 4\% of the total flux is concentrated in the
         core of the object (at 4.8 GHz) but they did not find any extended
         structures. Our A--Array radio map does not show any extent
         either, not 
         even on a tapered map. However, on an additional map taken with
         the VLA in the B--configuration, we detect diffuse emission
	 with an extent of nearly 20 kpc that was totally missed
         by our A--Array observations. Only 12\,\% of the B--Array
         flux was previously detected with the
         A--Array. Interestingly, the slight elongation 
         on the A--Array images is roughly east--west, but the B--Array
         image is elongated north--south.
 	 
	 {\bf PG0052+251}. For this source, Bennert et al. (2002)
         reported an extended structure in their HST [\ion{O}{iii}]
         emission-line map. This image is reproduced as a contour plot
         in Fig.~2. The corresponding radio map is also shown, but
         there is no elongation in the direction of the optical line
         emission component. The extension to the south--west
         on a 1.4 GHz map mentioned by Kukula et
         al. (\cite{kukula98}) in agreement with the optical image is
         not confirmed in our deep
         4.8 GHz map. As Kukula et al. (\cite{kukula98}) noted, this
         structure is possibly not real. Thus, there is no correlation between
         [\ion{O}{iii}] emission and radio emission for this
         separated knot.

         {\bf PG0157+001}. This triple radio source is clearly related
         to the emission--line structures shown by Bennert
         et al. (2002) (Fig.~3). The western knot of radio emission
         seems to represent the termination point of a jet, which is
         stopped in the interstellar medium. This is strengthened by a
         bow--shock 
         like structure of swept--up material in the emission--line
         image which is very 
         similar to what is seen in Mrk 573 (Pogge \& de Robertis
         \cite{pogge95}, Falcke et al. \cite{falcke98}, Ferruit et
         al. \cite{ferruit99}). As for Mrk 573, the interaction of the
         radio jet with the NLR gas producing the bow--shock is confirmed
         by optical spectroscopy (Leipski \& Bennert
         \cite{leipski06}). There is no corresponding association 
         of the optical 
         emission line structures at the position of the eastern radio
         knot. In fact, 
         the emission line gas is distributed to the south of this
         knot.\\\indent
         The eastern radio knot was missed by the
         snapshot images presented by Kellermann et
         al. (\cite{kellermann89}). The overall structure of the
         triple radio source resembles that of the large scale
         [\ion{O}{iii}] emission shown by Stockton \& MacKenty
         (\cite{stockton87}).
         A VLA B--Array map shows no additional components in this source.

  	  \begin{figure*}
         \centering
         \includegraphics[width=6cm,angle=-90]{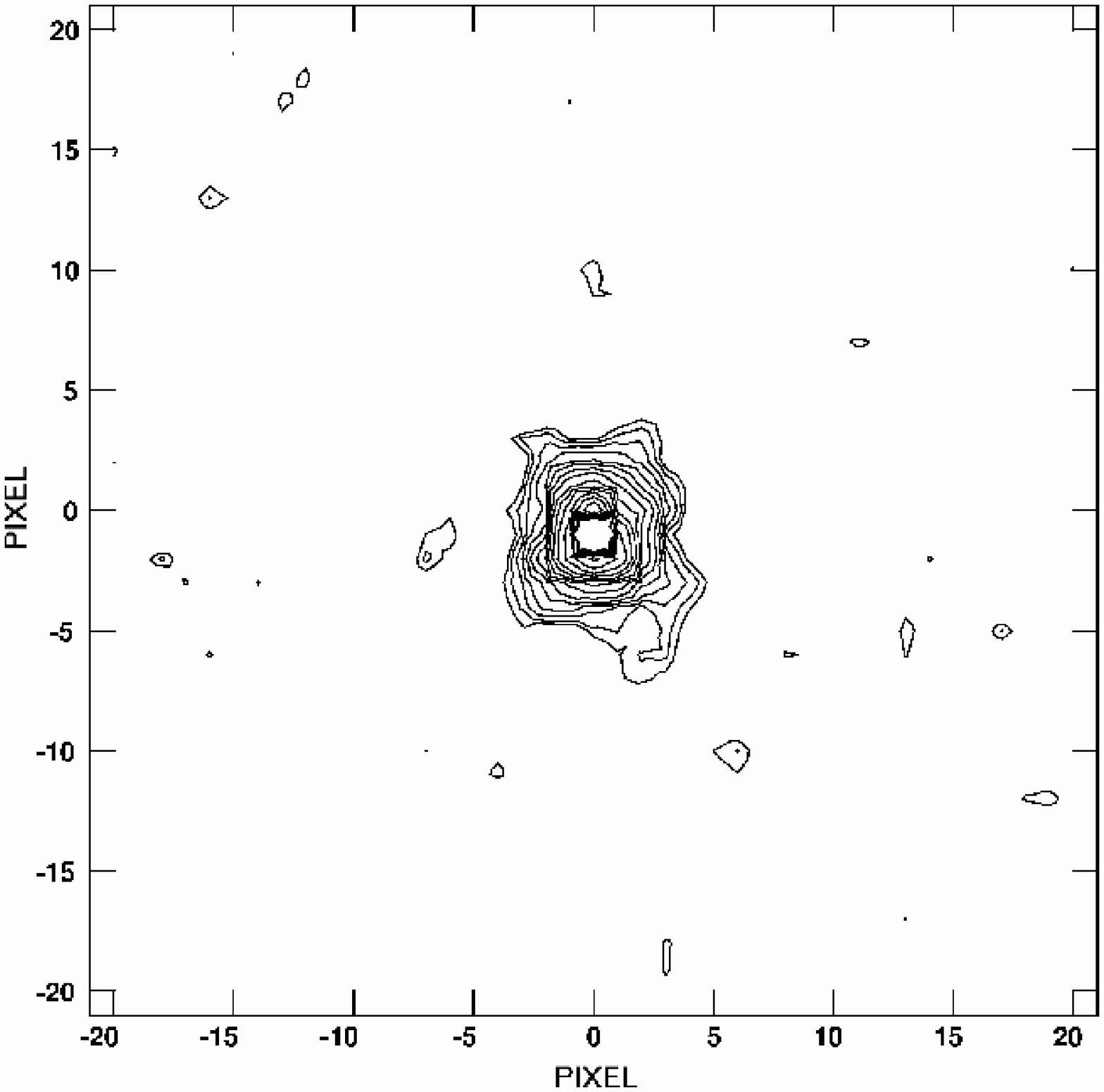}
         \includegraphics[width=6cm,angle=-90]{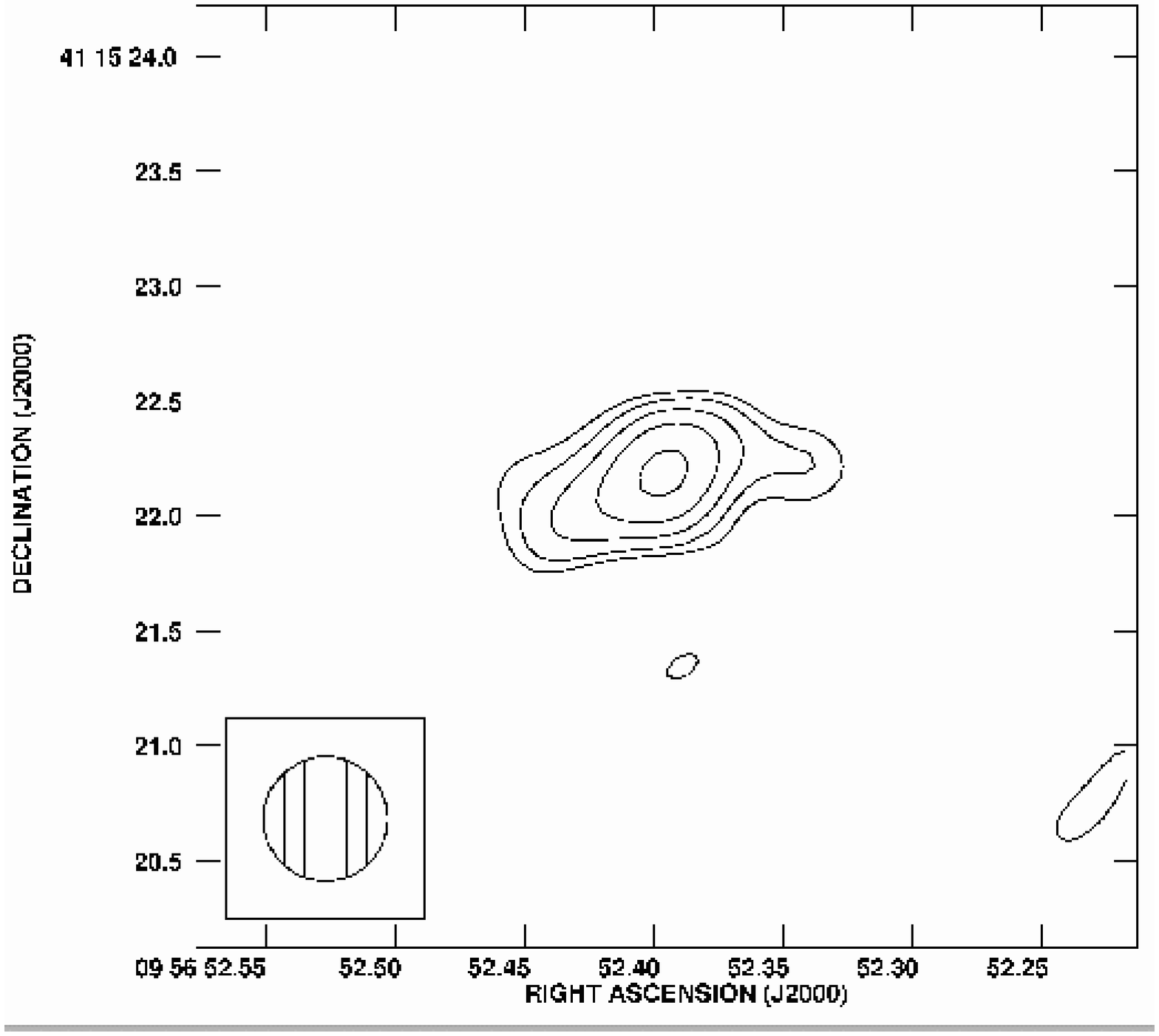}\\\hspace*{-0.5cm}
         \caption[]{\label{pg0953}Same as in Fig.\,2 for
         PG0953+414. The VLA map is at 4.8 GHz with a
         $0.^{\hspace*{-0.1cm}\prime\prime}54$ beam.}
         \end{figure*}
       \begin{figure*}
         \centering
         \includegraphics[width=6cm,angle=-90]{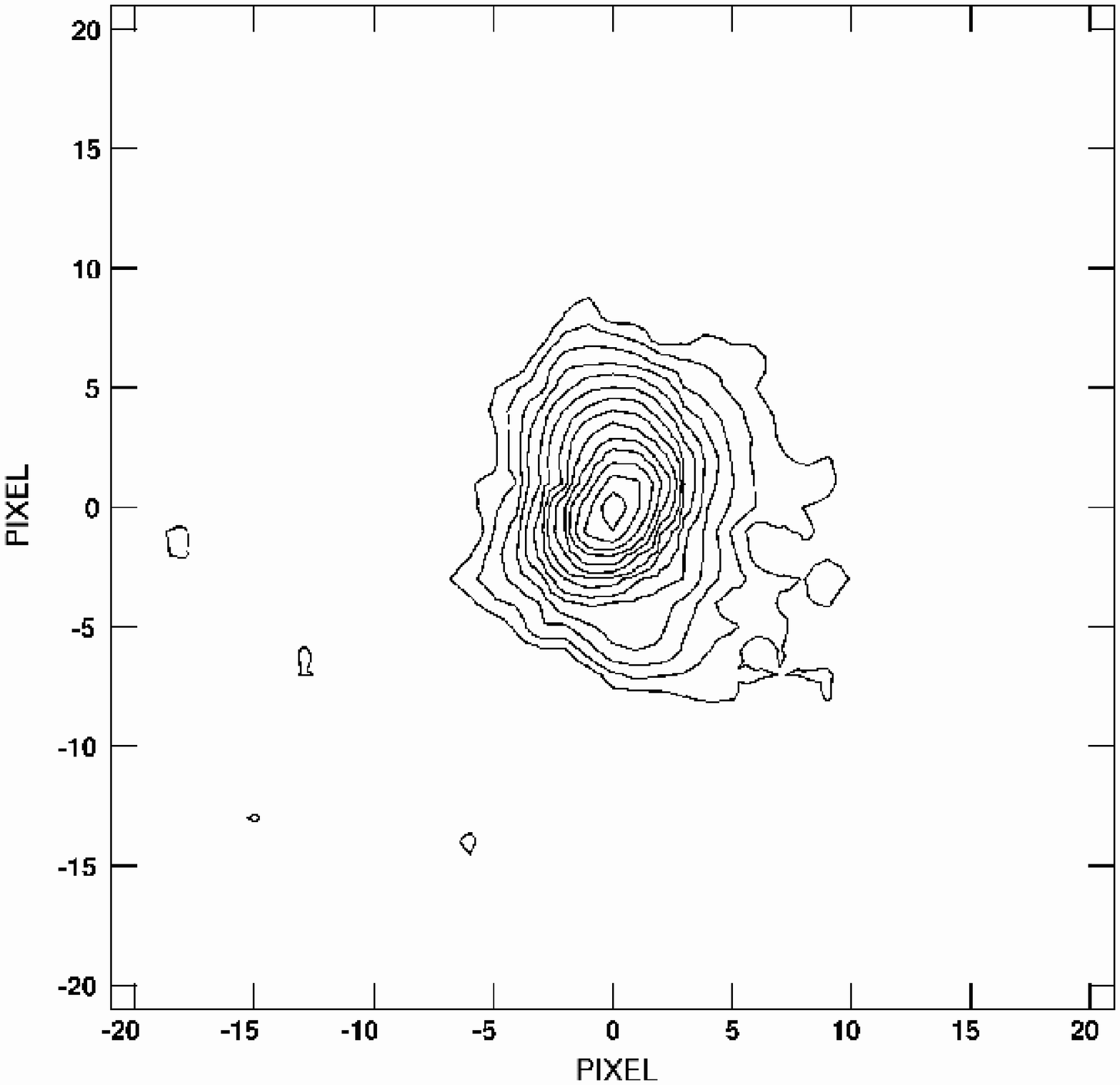}
         \includegraphics[width=6cm,angle=-90]{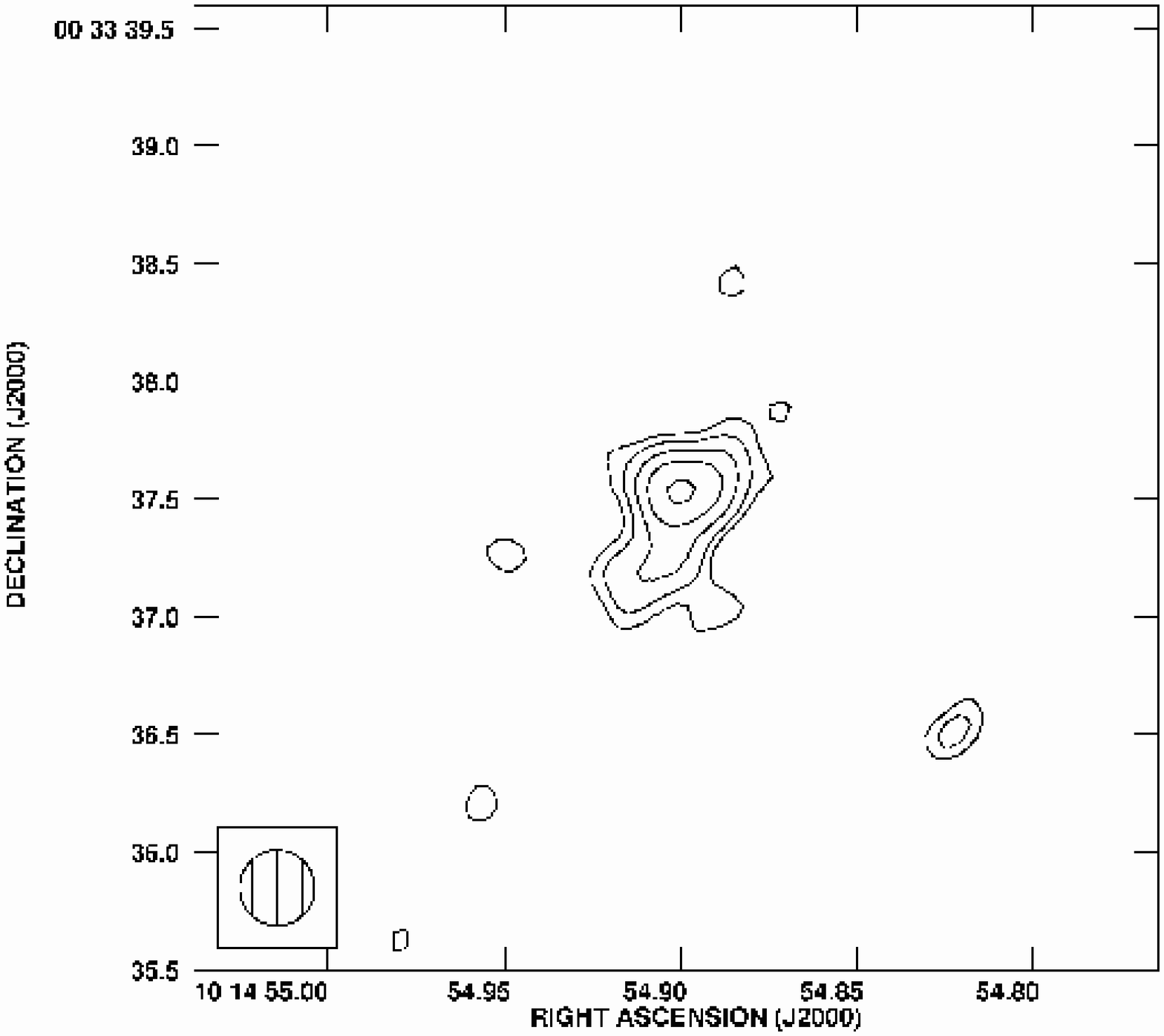}\\\hspace*{-0.5cm}
         \caption[]{\label{pg1012}Same as in Fig.\,2 for
         PG1012+008. The VLA map is at 8.4 GHz with a
         $0.^{\hspace*{-0.1cm}\prime\prime}32$ beam.}
         \end{figure*}
         {\bf PG0953+414}. This quasar is elongated in the
         east--west direction in the radio regime (Fig.\,4, right). In the
         corresponding [\ion{O}{iii}] emission--line image there is no
         significant extent, especially not in the direction of the
         radio extension (Fig.\,4, left).
         The radio structure seen with the A--Array
         does not appear on a B--Array map where the source is
         unresolved at 4.8 GHz. Thus, a radio jet seems
         to be directed to the east--west
         causing extended but collimated radio
         emission on scales of a few kpc. This source was not detected by 
         Kellermann et al. (\cite{kellermann89}).

         {\bf PG1012+008}. This quasar is part of an interacting system
         (Bahcall et al. \cite{bahcall97}). The interstellar
         medium is perturbed and it is difficult to say whether the
         structures are caused by the merger or by the radio--jet
         itself. The radio source consists of a nuclear source
         with a jet to the south (see Fig.\,2, right).
         The jet seems to bend to the east
         as indicated by comparing the inner and outer
         isophotes. The same "bended" structure is apparent in the
         optical image where the inner isophotes display the shape of
         the radio source (Fig.\,2, left). This indicates that
         the radio jet and the NLR gas are intimately
         connected. However, it is not clear what effect the galaxy
         interaction has 
         on the apparent structures, especially on the "bending".\\\indent
         A corresponding B--Array map resolves no prominent additional 
         structures, except one possible short extension to the north--west in 
         the opposite direction of the jet detected in the A--Array.

         {\bf PG1049$-$005}. This source appears pointlike at both wavelengths
         (Fig.\,6), 
         but the HST image is resolved due to the better resolution
         (Bennert et al. \cite{bennert02}).
	 For the radio extent we can only give an upper limit. Note
         that the elongations in the HST image (Fig.\,6, left) to
         the north--east and south--west and the minor elongations to
         the south--east and north--west are due to the secondary
         mirror supporting structure and, therefore, are artificial.

         {\bf PG1307+085}. At both (radio and optical) wavelengths, a compact
         component can be seen 
         as well as an elongation to the south--west
         (Fig.\,7). While the radio map displays a secondary
         component in the south--west that possibly is part (or the
         termination point) of a radio jet, the HST image shows
         diffuse emission in the same region. While these results
         are consistent with a radio--NLR interaction, they are by no
         means conclusive.  
       \begin{figure*}
         \centering
         \includegraphics[width=6cm,angle=-90]{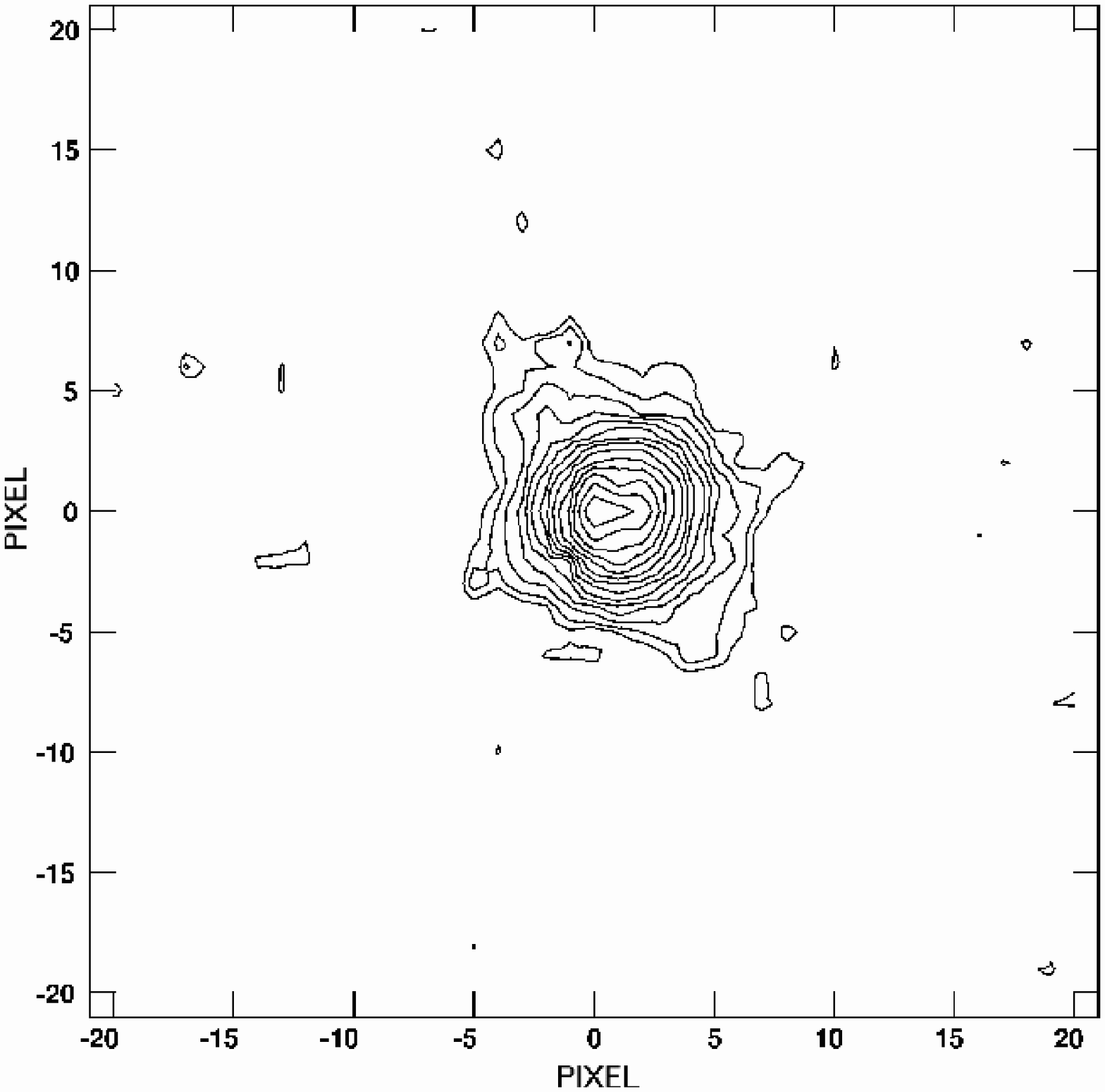}
         \includegraphics[width=6cm,angle=-90]{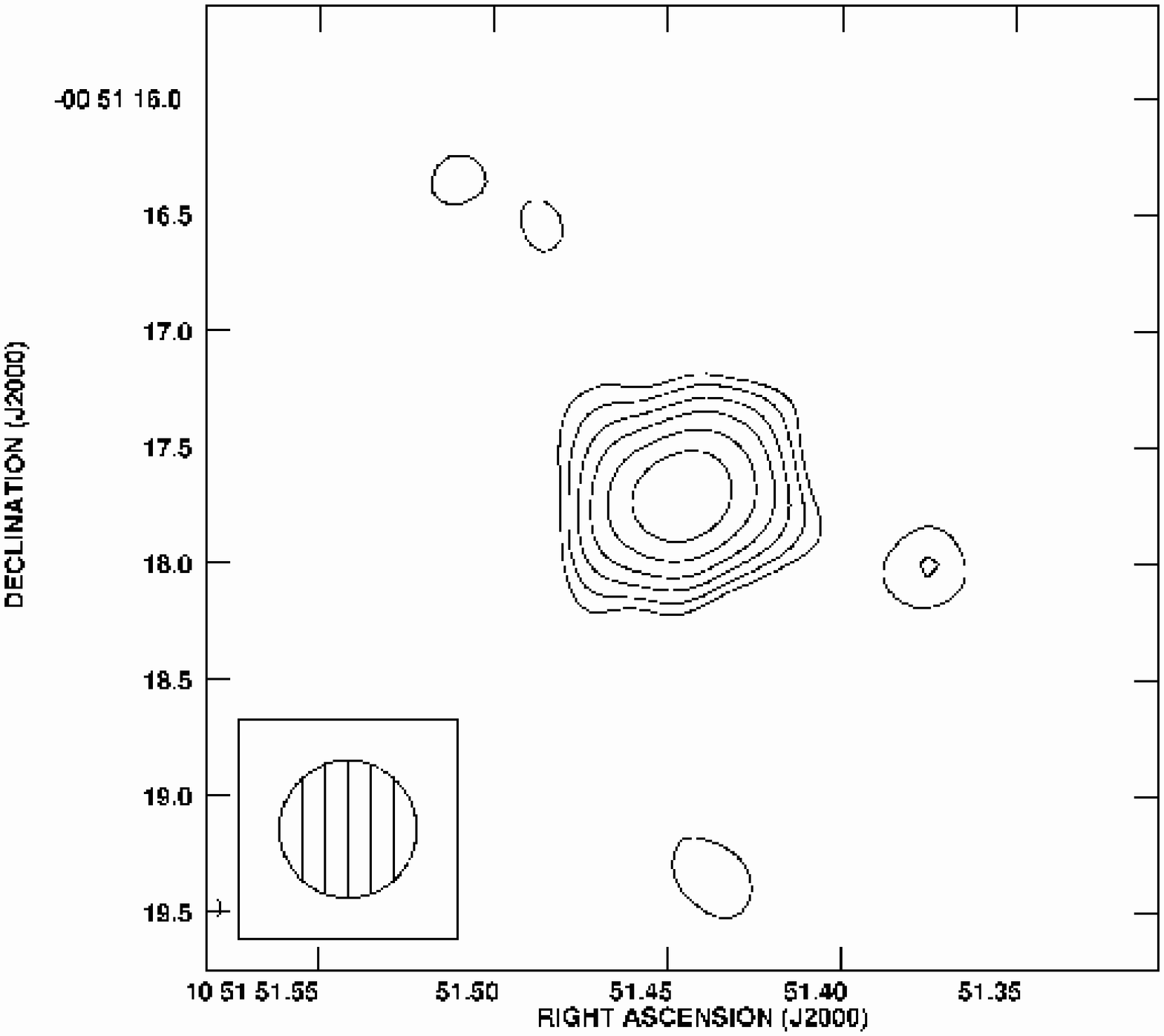}\\\hspace*{-0.5cm}
         \caption[]{\label{pg1049}Same as in Fig.\,2 for
         PG1049-005. The VLA map is at 4.8 GHz with a
         $0.^{\hspace*{-0.1cm}\prime\prime}59$ beam.}
         \end{figure*}
         \begin{figure*}
         \centering
         \includegraphics[width=6cm,angle=-90]{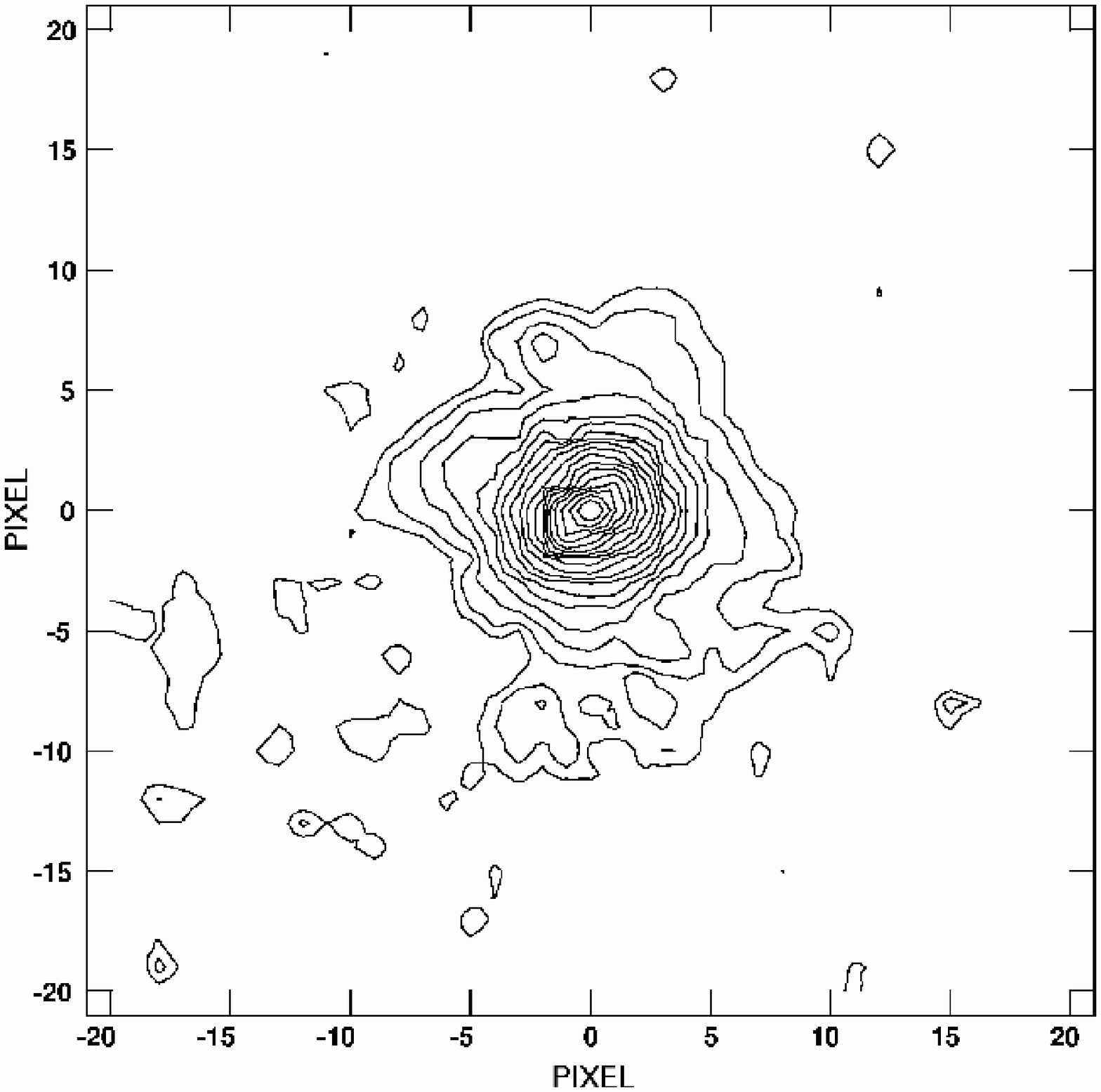}
         \includegraphics[width=6cm,angle=-90]{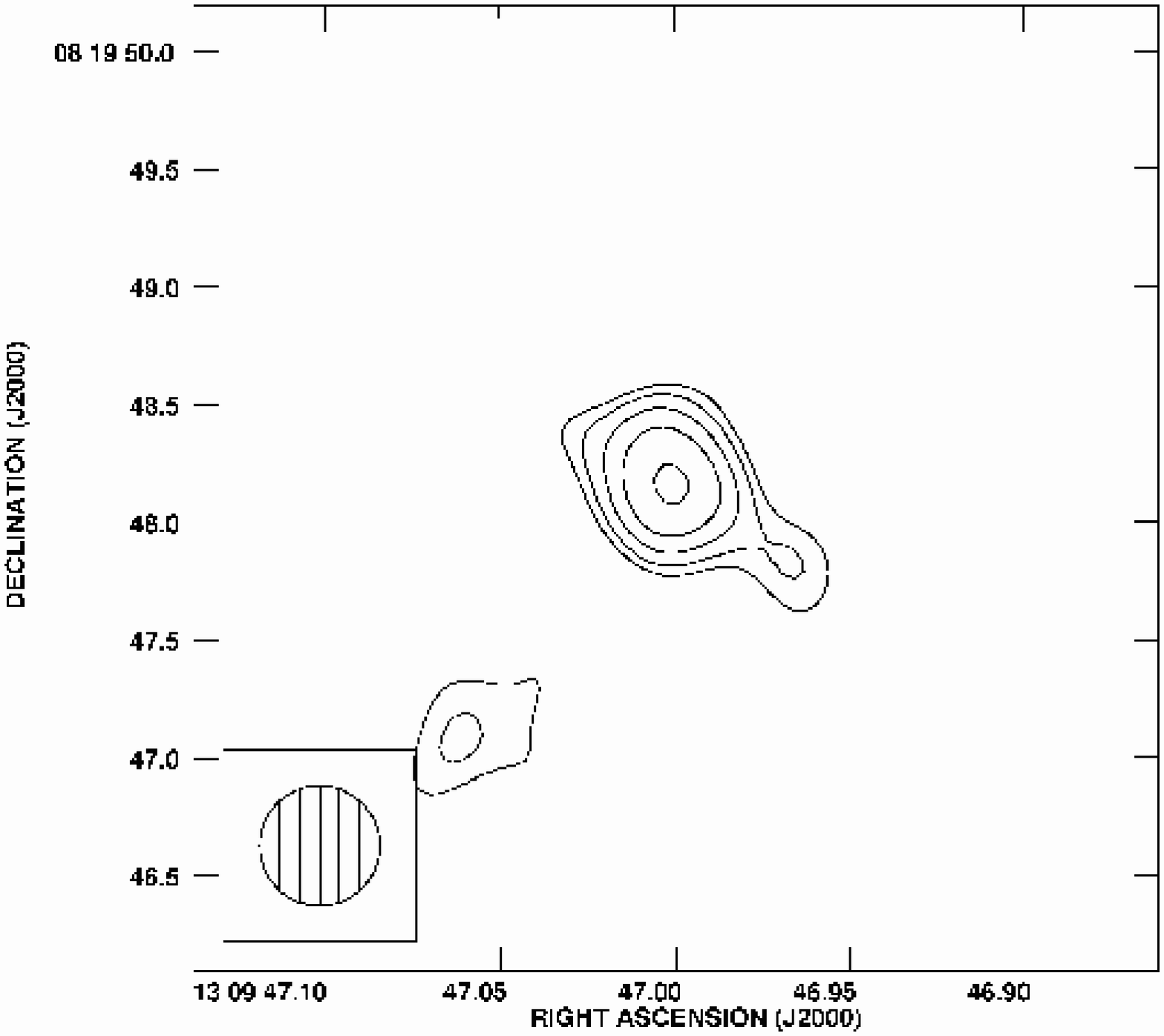}\\\hspace*{-0.5cm}
         \caption[]{\label{pg1307}Same as in Fig.\,2 for
         PG1307+085. The VLA map is at 4.8 GHz with a
         $0.^{\hspace*{-0.1cm}\prime\prime}51$ beam.}
         \end{figure*}
 \subsubsection{Seven RQQs without HST NLR images}
  For the following seven objects (the middle part of Table \ref{sample}
  and \ref{sample_results}), no optical emission--line images are
  available. Their redshifts and luminosities lie in between the redshifts and
  luminosities of the RQQs mentioned above and the Seyfert galaxies.
  Thus, these seven objects are very interesting as they connect
  Seyferts and RQQs. If there is indeed a close connection between
  {\sl all} radio--quiet AGN, it shall be reflected by the radio
  structures in these objects.

{\bf PG0003+199}. A compact, unresolved radio source at 4.8 GHz
(Fig.\,8, left).

{\bf PG1119+120}. Three different components are visible in this
                source: A compact component in the 
                south--west and at the centre as well as a one--sided jet
                to the north--east (Fig.\,9). This jet is nearly two 
                arcseconds long before it sharply bends to the
                north--west. The whole structure is embedded  
                in a more diffuse emission, especially to the north of
                the south--western blob (Fig.\,9, right). Leipski
                \& Bennert (\cite{leipski06}) have shown that extended
                [\ion{O}{iii}] emission exists in this source and that
                the emission--line ratios suggest an interaction of
                the radio jet with the NLR gas.   

{\bf PG1149$-$110}. A triple radio source extending
                  $1.^{\hspace*{-0.1cm}\prime\prime}5$  
                  in roughly east--west direction (Fig.\,10,
                  left). Although the western component seems to become steadily
                  weaker it has a peak that
                  can be seen with different contour  
                  spacings. Neither Kellermann et
                  al. (\cite{kellermann89}) nor Kukula et
                  al. (\cite{kukula98}) were able to detect any 
                  components in addition to an unresolved core. 

{\bf PG1351+640}. This source is known to be strongly variable
                (Barvainis \& Antonucci \cite{barvainis89})  
                and has a VLBA core (Blundell \& Beasley
                \cite{blundell98}, Ulvestad et
                al. \cite{ulvestad05}). However, on arcsecond scales
                we detect three aligned components  
                with surrounding diffuse emission (Fig.\,10, right). 
               Kellermann et al. (\cite{kellermann89,kellermann94}) have only 
               securely detected the central component.

{\bf PG1534+580}. Unresolved like PG0003+199 (Fig.\,8, right).

{\bf PG1612+261}. An one--sided jet heads
                $1.^{\hspace*{-0.1cm}\prime\prime}5$ towards the
                south--west  
                (Fig.\,11, left). As in the case 
                of , three peaks can be
                identified in this source with adequate contour spacings.  
                While one peak lies very close in the north--east of
                the central peak, another peak appears at the
                tip of the jet. Nevertheless, this source holds its
                one--sidedness. The R value 
                of PG1612+261 is remarkably high (2.81,
                Kellermann et al. 1989) for a radio--quiet quasar and
                comparison with flux values given by Kellermann et
                al. (\cite{kellermann89}) indicates variability. 
                On an additional B--Array map there is no further
                structure detectable and only 14\% of the B--Array 
                flux is missed by the A--Array observations.  

{\bf PG2130+099}. Another aligned triple radio source, but with well
                separated and clearly detected components 
                (Fig\,11, right). The overall extent is nearly
                $2\farcs5$, orientated north--west/south--east. Especially the
                component to the north--west displays a  
                remarkable sub--structure which is elongated
                perpendicular to the jet--direction, possibly
                indicating a transverse shock.
		
 \begin{figure*}
         \centering
         \includegraphics[width=6cm,angle=-90]{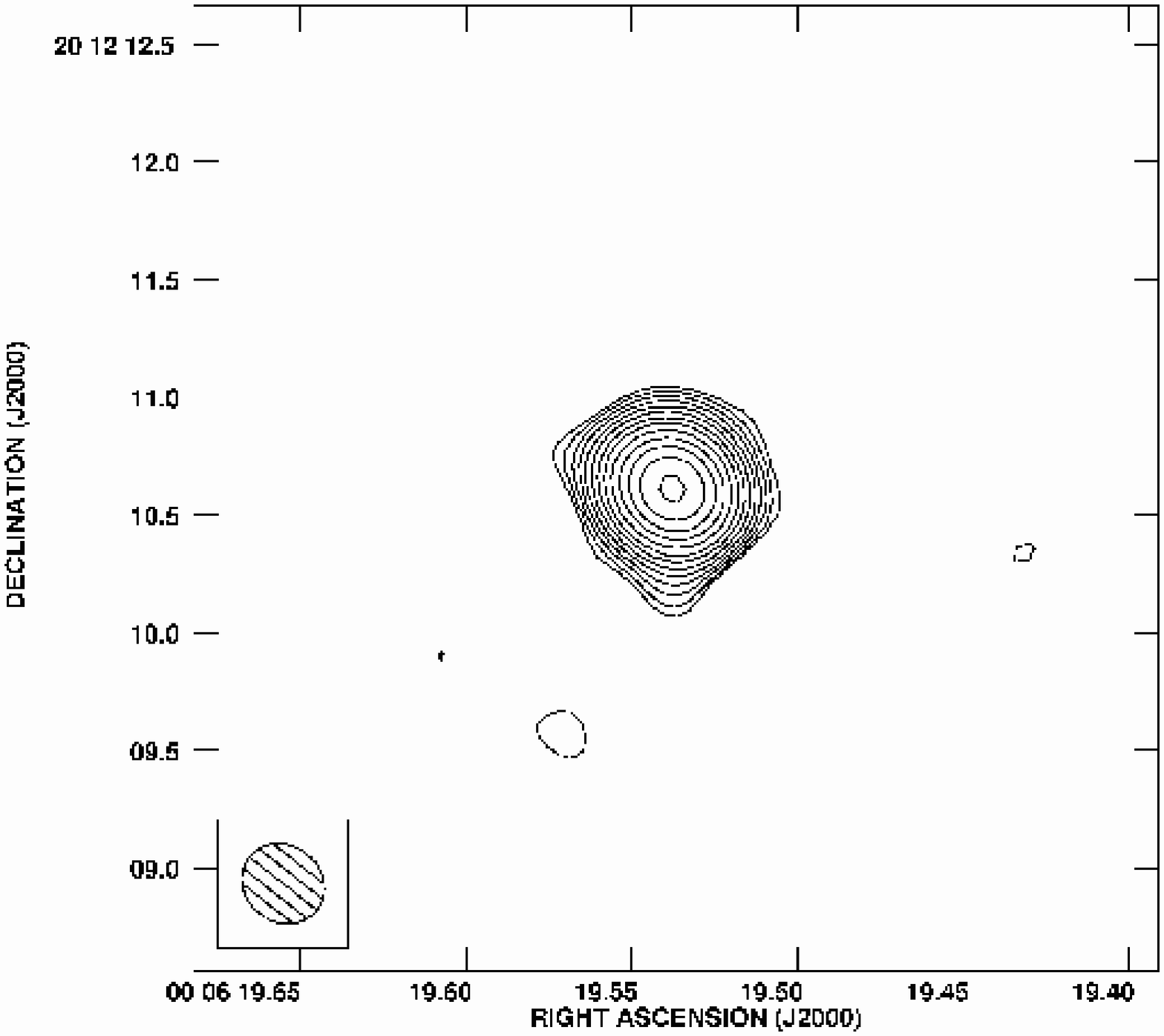}
         \includegraphics[width=6cm,angle=-90]{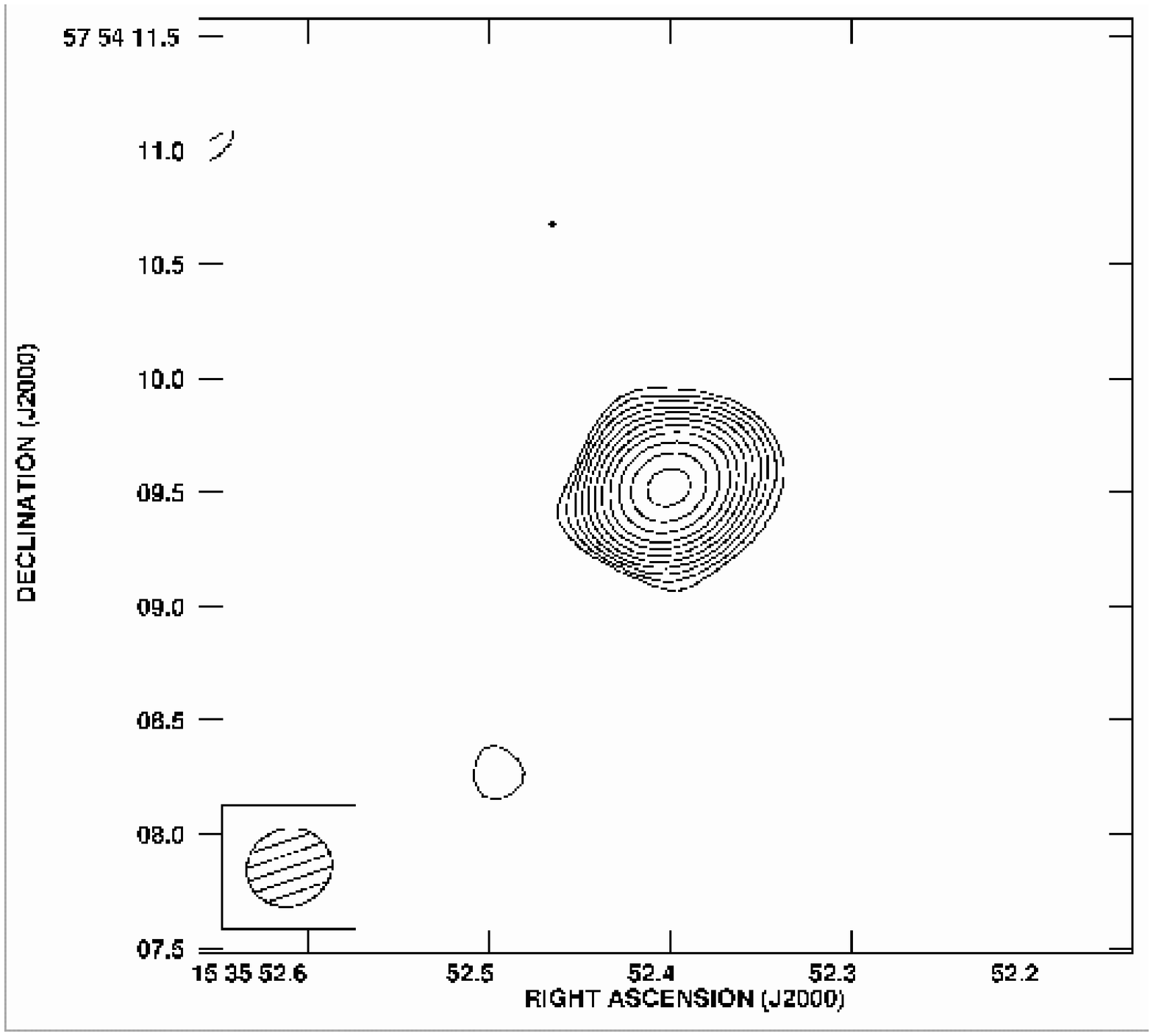}\\\hspace*{-0.5cm}
         \caption[]{\label{pg0003}VLA images of
         PG0003+199 (left) and PG1534+580 (right). Both images are
         $4^{\prime\prime}\times4^{\prime\prime}$ wide. {\sl Left}: Uniform--weighted VLA
         A--Array map at 4.8 GHz with a $0.^{\hspace*{-0.1cm}\prime\prime}37\times0.^{\hspace*{-0.1cm}\prime\prime}32$ beam
         {\sl Right}:
         Uniform--weighted VLA
         A--Array map at 4.8 GHz with a $0.^{\hspace*{-0.1cm}\prime\prime}38\times0.^{\hspace*{-0.1cm}\prime\prime}34$ beam.}
         \end{figure*}
\begin{figure*}
         \centering
         \includegraphics[width=6cm,angle=-90]{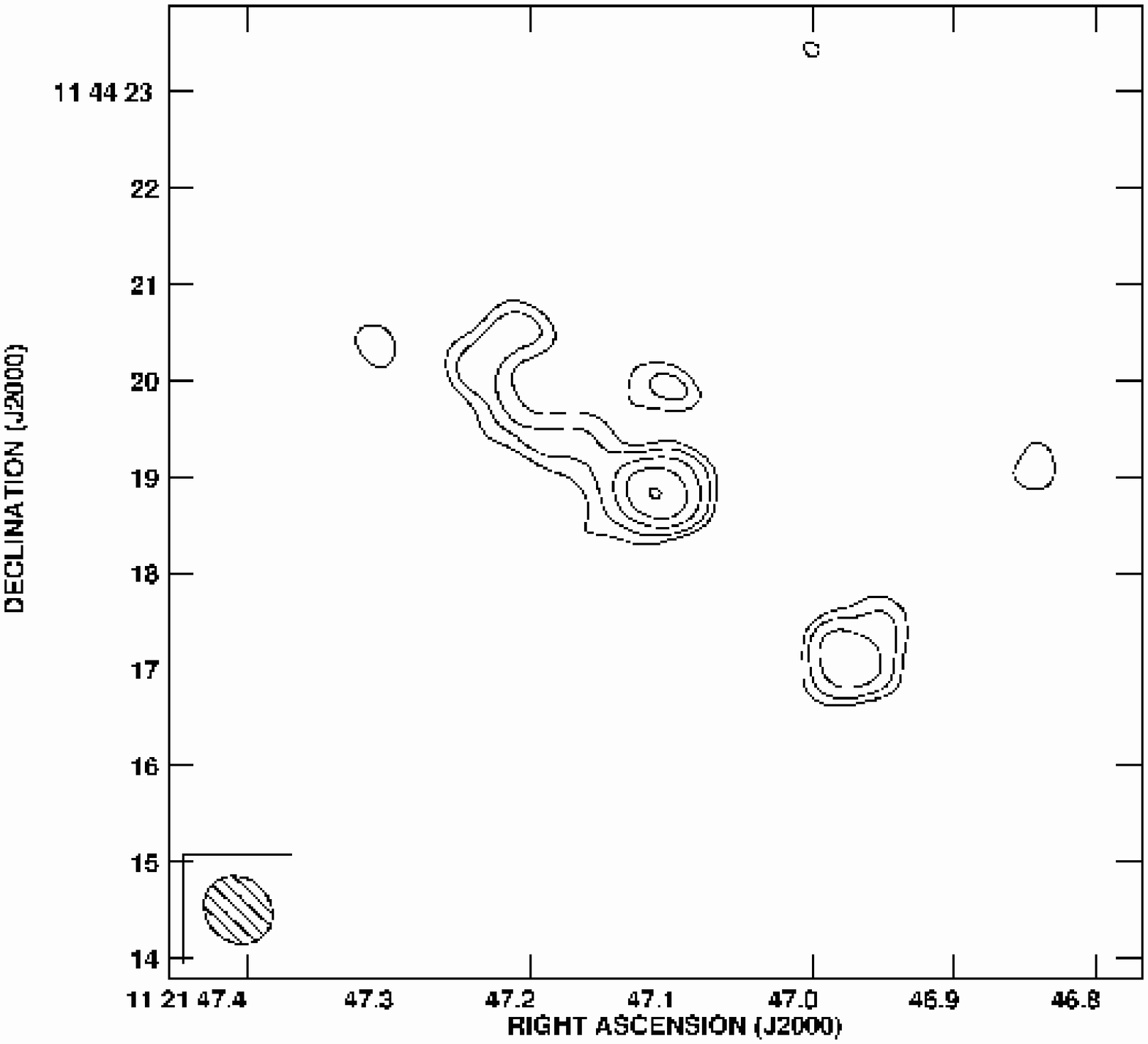}
         \includegraphics[width=6cm,angle=-90]{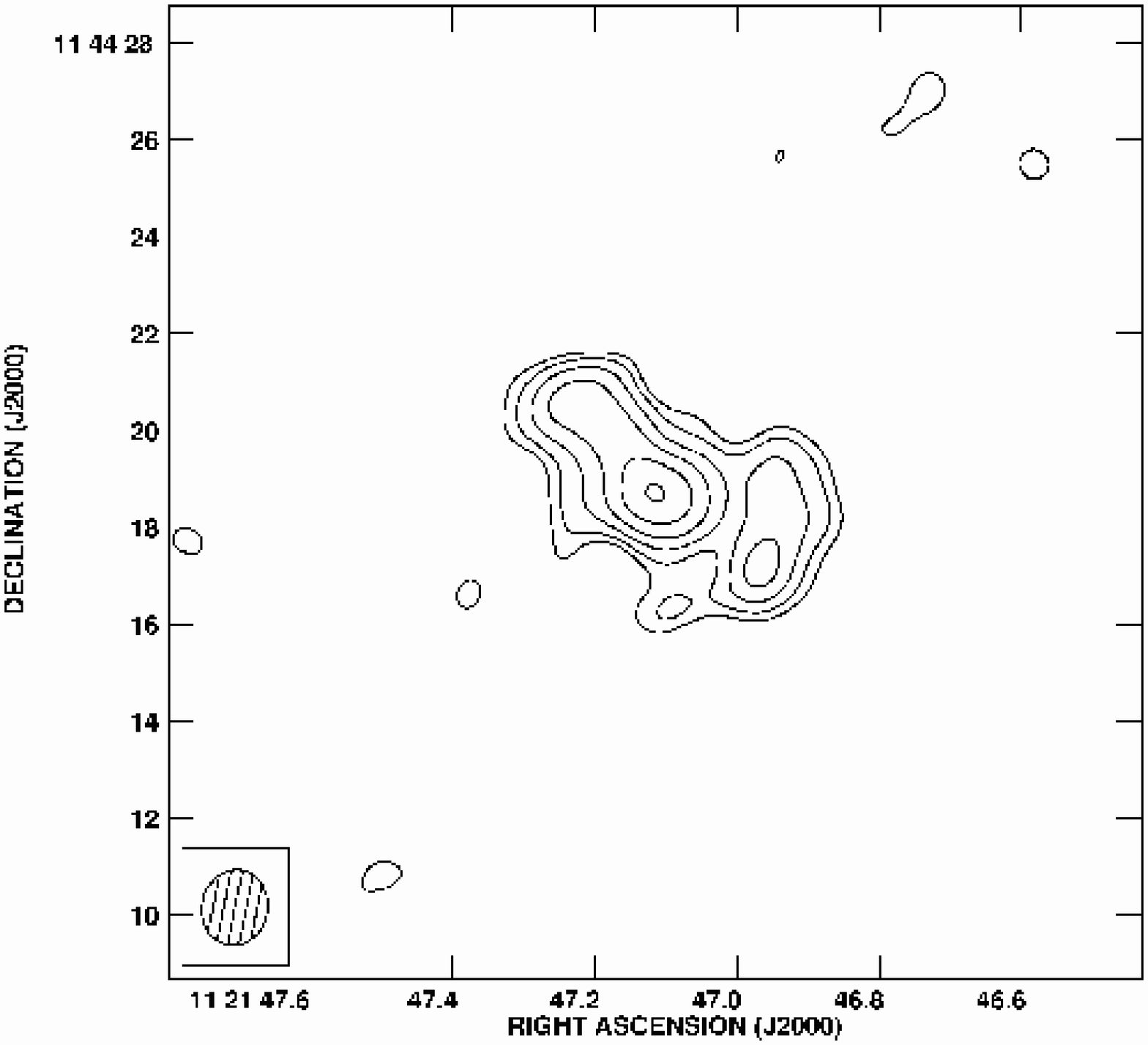}\\\hspace*{-0.5cm}
         \caption[]{\label{pg1119}VLA images of
         PG1119+120. {\sl Left}:
         $10^{\prime\prime}\times10^{\prime\prime}$ wide
         image. Natural--weighted A--Array map at 4.8 GHz with a UV taper 300
         k$\lambda$ and a
         $0.^{\hspace*{-0.1cm}\prime\prime}75\times0.^{\hspace*{-0.1cm}\prime\prime}66$
         beam.
	 {\sl Right}: $20^{\prime\prime}\times20^{\prime\prime}$ wide
         image. Natural--weighted VLA
         B--Array map at 4.8 GHz with a $1.^{\hspace*{-0.1cm}\prime\prime}6\times1.^{\hspace*{-0.1cm}\prime\prime}4$ beam.}
         \end{figure*}

\section{Discussion}
\subsection{Structure of the radio emission}
a) Sizes and luminosities\\*[2mm]\noindent
To investigate the radio emission of radio--quiet
  quasars we first compare their sizes and luminosities with
  literature data before studying their morphologies in
  detail. \\\indent
  In luminosity, there is almost no overlap between the higher 
  redshift RQQ sample and local ($v < {\rm 5000~km\,s^{-1}}$) Seyfert
  samples (e.g. Morganti et al. \cite{morganti99}, Ulvestad \& Ho
  \cite{ulvestad01}, Ho \& 
  Ulvestad \cite{ho01}). However, the very bright end of the (rather
  powerful) 12 $\mu$m + CfA sample extends well inside (but does not
  exceed) the 
  luminosities of the higher redshift RQQs (Rush et
  al. \cite{rush96}). On the other hand, the lower redshift RQQ sample
  connects the almost distinct local Seyfert and high--$z$ RQQ
  samples in terms of luminosity, but they hardly
  exceed the very brightest 12 $\mu$m + CfA sources. Therefore, the
  higher redshift sample significantly increases the luminosity
  baseline for radio--quiet AGN, while the lower redshift sample aids to
  bridge Seyferts and quasars.\\\indent
  Considering the sizes, a very similar
  situation is achieved. But this time the size of even the largest
  RQQ in our study hardly exceeds the size of the largest Seyferts
  (e.g. Morganti et al. \cite{morganti99}, Ulvestad \& Ho
  \cite{ulvestad01}). However, the size of a RQQ with a given
  morphology is in fact larger than that of Seyferts of the {\sl same}
  morphological type (see discussion below).
  Additionally, at higher luminosities
  the situation may be much more complex due to the possibility of AGN
  to harbour large 
  scale low flux density emission (Fig\,\ref{pg0026}).
  Moreover, at the high luminosity end of our samples we start to
  enter the range in size and luminosity measured for the radio
  galaxies with the very lowest luminosities (de Ruiter et
  al. \cite{deruiter90}, Neeser et al. \cite{neeser95}).\\\indent 
  In the well known L$_{\rm radio}$--L$_{\rm [\ion{O}{iii}]}$ correlation
  for AGN (e.g. Miller et al. \cite{miller93}, Falcke et
  al. \cite{falcke95}, Xu et
  al. \cite{xu99}), the radio--quiet population is clearly
  distinguished from the radio--loud population. While both
  populations share roughly the same range in L$_{\rm [\ion{O}{iii}]}$,
  they are separated in L$_{\rm radio}$ with only very little overlap. 
  Given that L$_{\rm [\ion{O}{iii}]}$ can
  be used as a measure for the total luminosity of the
  AGN\footnote{Although this is widely accepted, recent results
  provide evidence that this simple picture might not be true (Haas et
  al. \cite{haas05}).}, some radio--quiet objects with the very highest
  (total) luminosities have sizes comparable to the radio--loud objects
  with the very lowest (total) luminosities\footnote{This obviously
  reflects the distinction of radio--loud and radio--quiet objects via
  the R--parameter (e.g. Kellermann et
  al. \cite{kellermann89}).}. However, in this apparent
  continuity the overlap between our radio--quiet objects and the radio--loud
  population in the radio regime is sparse and has to be addressed in detail
  elsewhere. \\\noindent
  \newline
  b) Detailed morphological comparison\\*[2mm]\noindent
The radio--quiet quasars observed here can be roughly divided into four
groups on the basis of their radio morphology: \\
(i) quasars that are unresolved (or almost unresolved) and point--like
(e.g. PG0052+251, Fig.\,\ref{pg0052}),\\ 
(ii) objects which exhibit symmetric, double--sided structures that steadily
fade out (e.g. PG0953+414, Fig.\,\ref{pg0953}),\\
(iii) objects with clear sub--peaks (radio knots) on both side of
the nucleus (e.g. PG0157+001, Fig.\,\ref{pg0157}),\\
(iv) quasars with strong asymmetry towards one side of the nucleus
(e.g. PG1119+120, Fig.\,\ref{pg1119}).\\\indent
All these types of radio structure can also be found in Seyfert
galaxies (e.g. Ulvestad \& Wilson \cite{ulvestad89}, Kukula et
al. \cite{kukula95}, Kinney et al. \cite{kinney00}, this paper).
As the Seyfert galaxies presented here were
selected on the basis of known extended radio emission (see \S\,2 and
Appendix A), we
do not find counterparts to quasars in group (i). However, many
Seyfert galaxies are known to be unresolved sources, even
with sub--arcsecond resolution: Comparing the radio
snapshot--surveys of Seyfert galaxies (e.g.  
Ulvestad et al. \cite{ulvestad81}, Schmitt et al. \cite{schmitt01}) and 
quasars (e.g. Miller et al. \cite{miller93}, Kellermann et al. 
\cite{kellermann94}, Kukula et al. \cite{kukula98}), compact and
unresolved objects are  
common to both object classes. But, studies using deep images with 
long integration times (e.g. Falcke et al. \cite{falcke98}, Blundell \& 
Rawlings \cite{blundell01}, this paper) have shown that sources
appearing pointlike in snapshot surveys need not necessarily to be
unresolved. Instead, low flux--density emission (e.g. 
PG0026+129 (Fig.\,\ref{pg0026}), PG1119+120
(Fig.\,\ref{pg1119}), Blundell \& Rawlings \cite{blundell01}) and
  additional components (e.g  
PG0157+001 (Fig.\,\ref{pg0157})) could be detected --
particularly when multi--array observations are used.\\\indent
\begin{figure*}
         \centering
         \includegraphics[width=6cm,angle=-90]{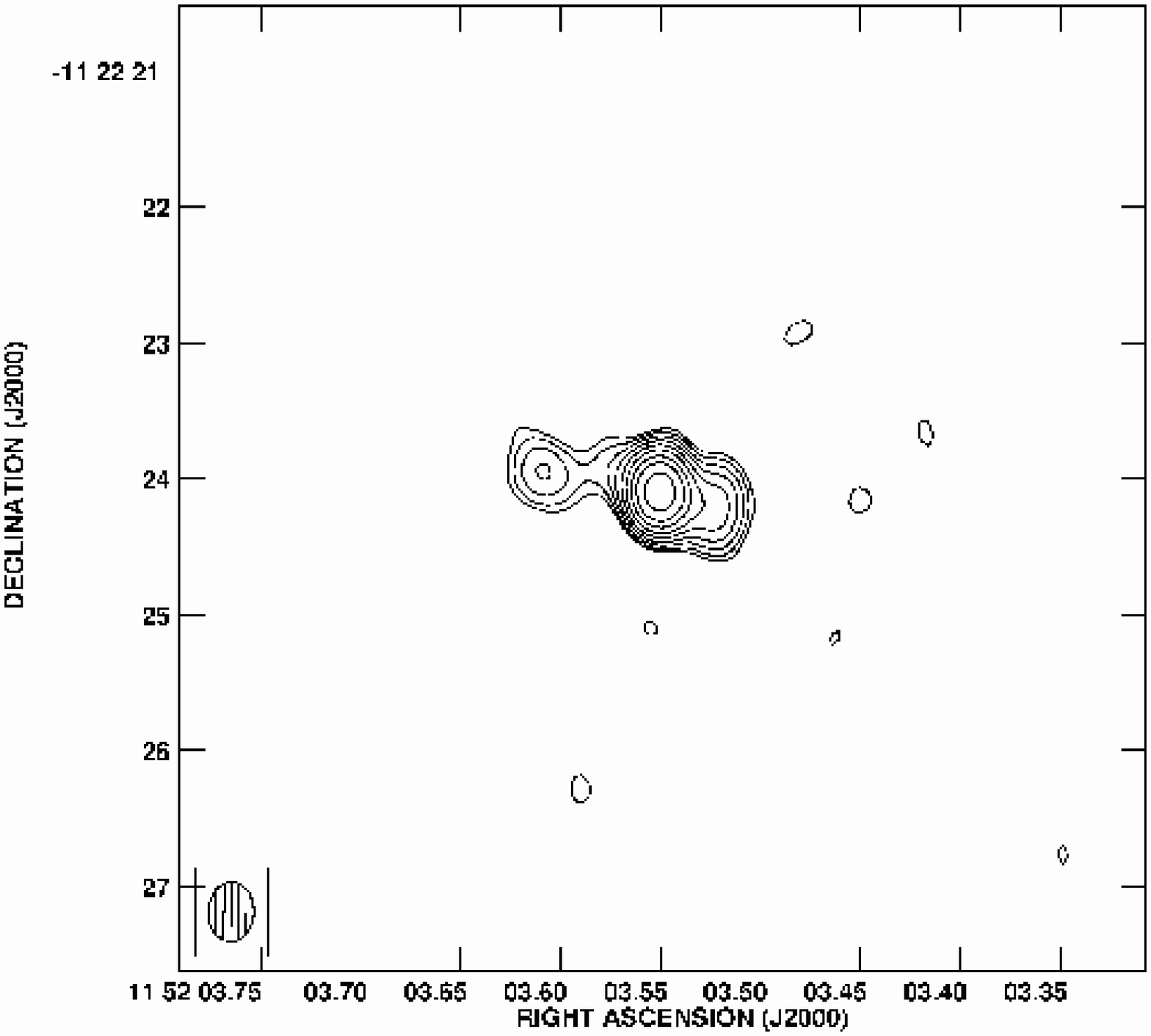}
         \includegraphics[width=6cm,angle=-90]{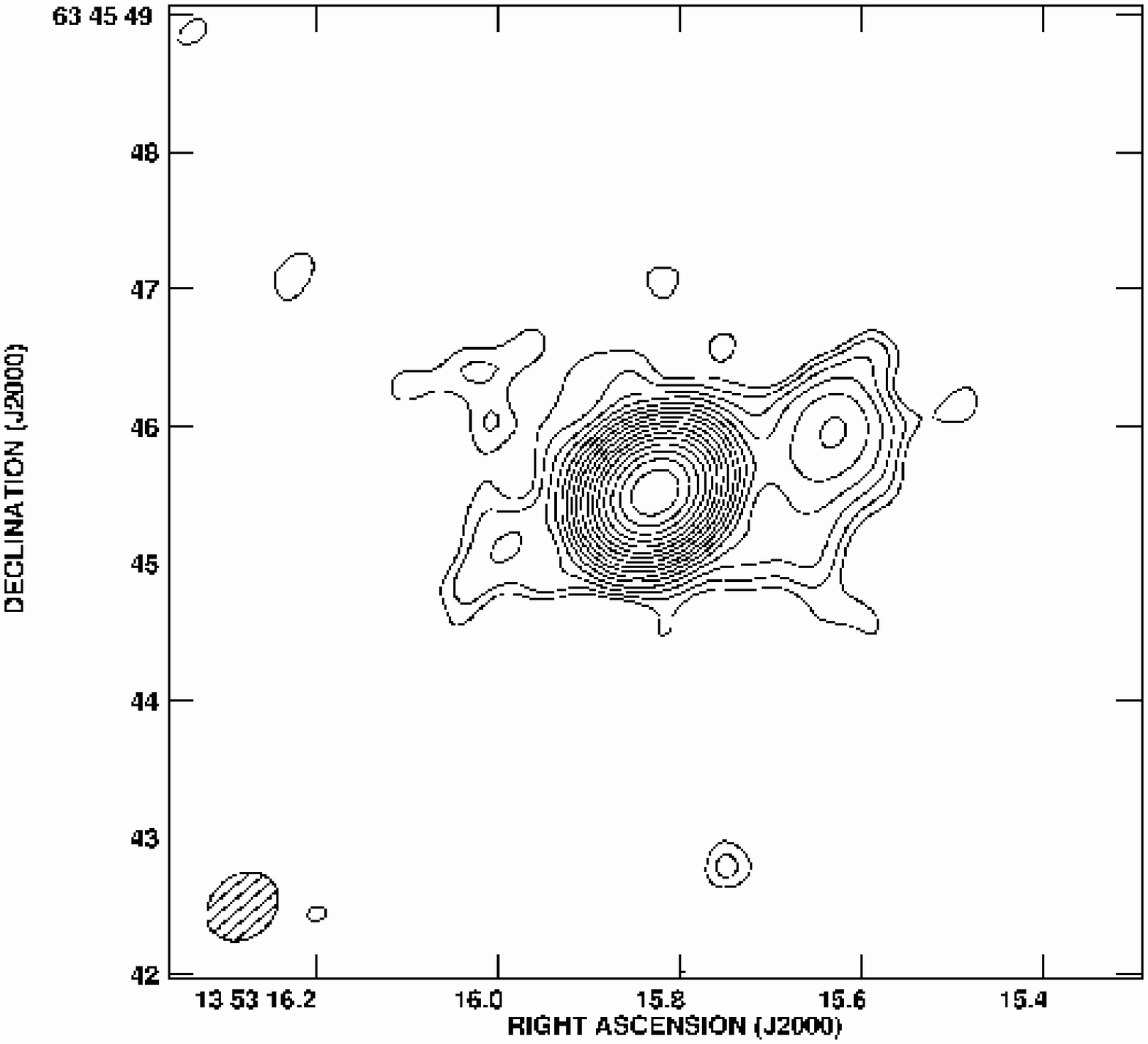}\\\hspace*{-0.5cm}
         \caption[]{\label{pg1149}VLA images of
          (left) and PG1351+640 (right). Both images are
         $7^{\prime\prime}\times7^{\prime\prime}$ wide. 
         {\sl Left}: Uniform--weighted A--Array map at 4.8 GHz with a  
         $0.^{\hspace*{-0.1cm}\prime\prime}44\times0.^{\hspace*{-0.1cm}\prime\prime}33$ beam.
         {\sl Right}: Natural--weighted VLA
         A--Array map at 4.8 GHz with a 
         $0.^{\hspace*{-0.1cm}\prime\prime}56\times0.^{\hspace*{-0.1cm}\prime\prime}45$}
         \end{figure*}               
\begin{figure*}
         \centering
         \includegraphics[width=6cm,angle=-90]{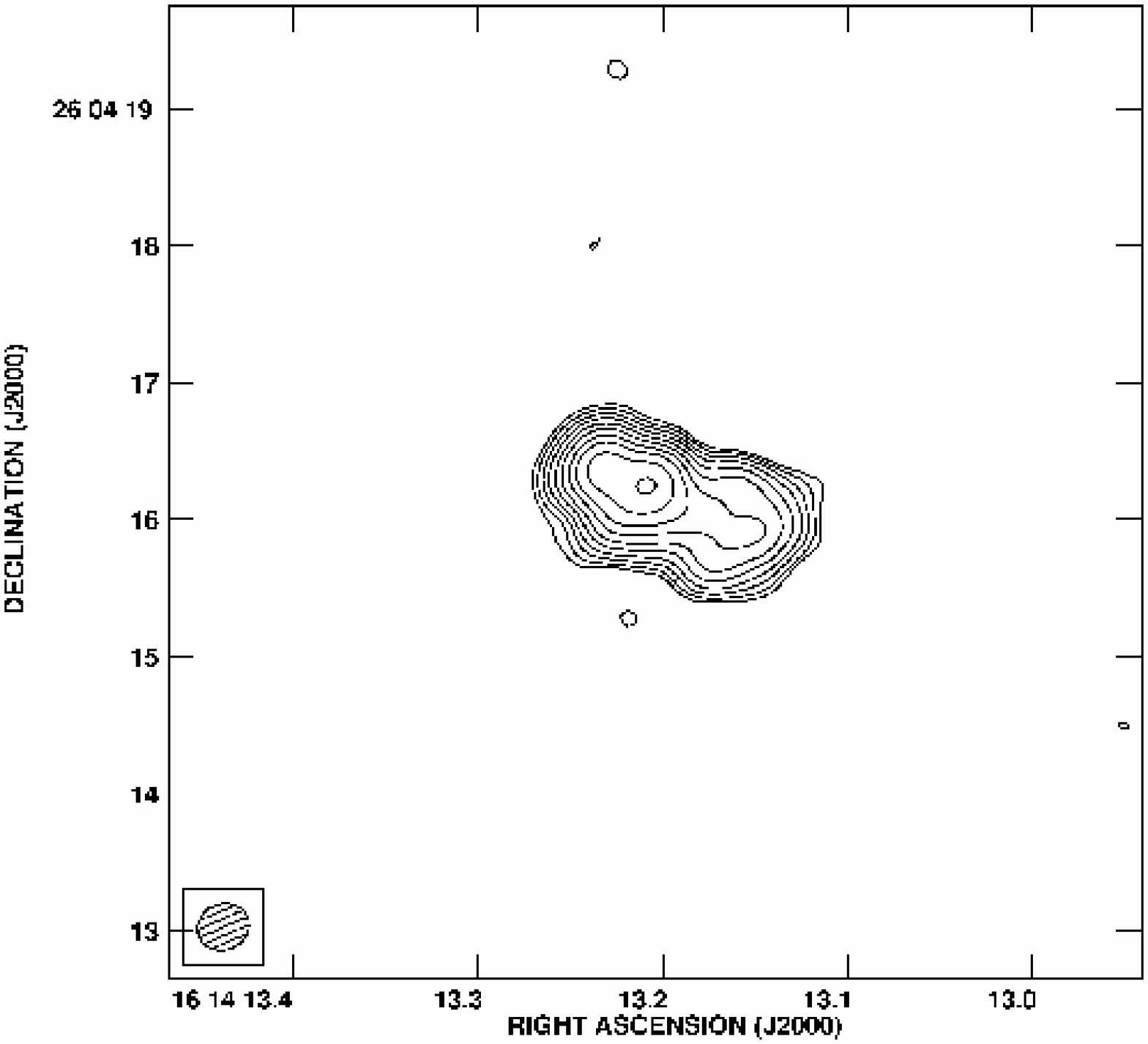}
         \includegraphics[width=6cm,angle=-90]{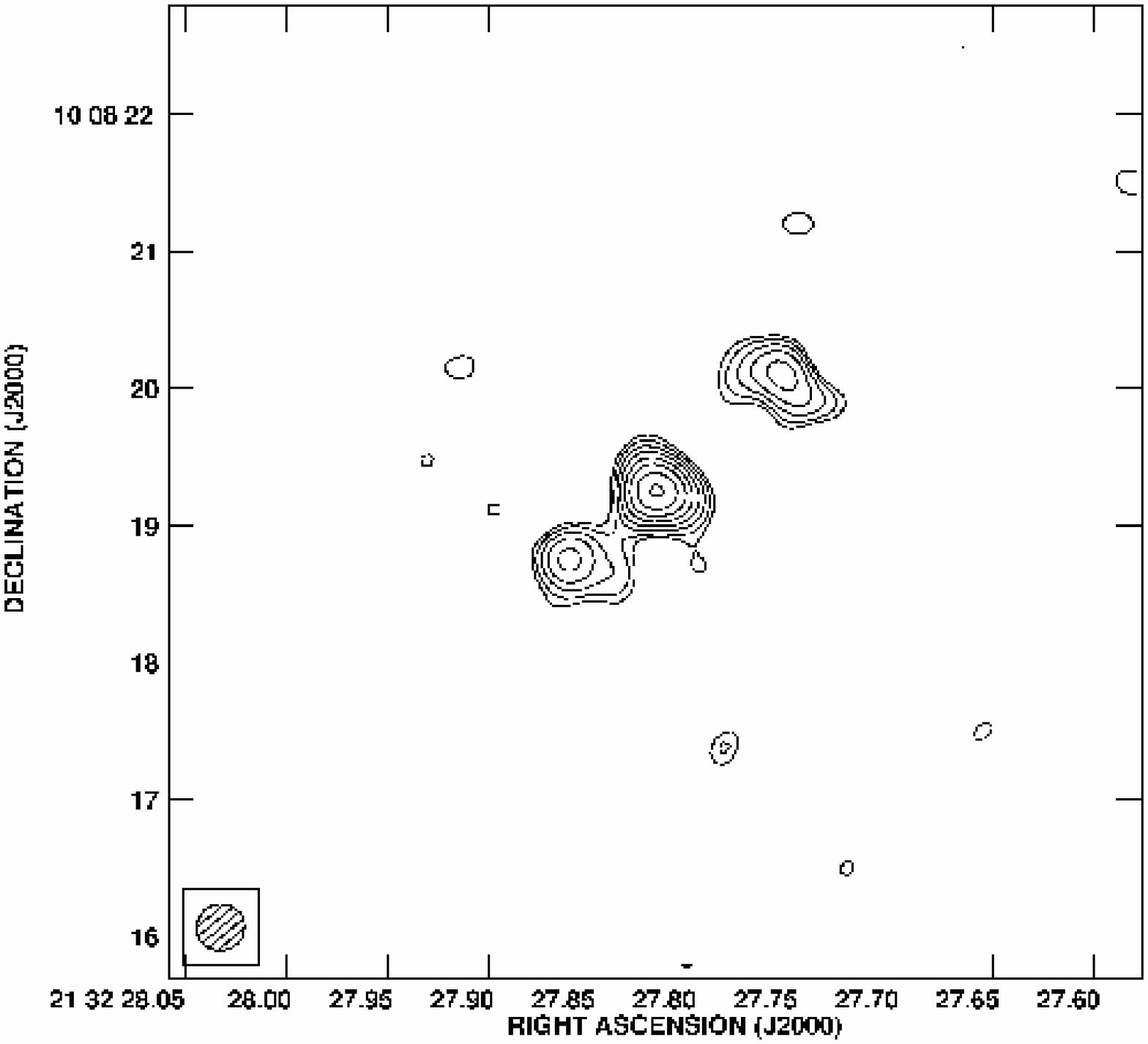}\\\hspace*{-0.5cm}
         \caption[]{\label{pg1612}VLA images of
         PG1612+261 (left) and PG2130+099 (right). All images are
         $7^{\prime\prime}\times7^{\prime\prime}$ wide. {\sl Left}: Uniform--weighted 
         A--Array map at 4.8 GHz with a  
         $0.^{\hspace*{-0.1cm}\prime\prime}38\times0.^{\hspace*{-0.1cm}\prime\prime}34$ 
         beam. 
{\sl Right}: Uniform--weighted VLA
         A--Array map at 4.8 GHz with a 
         $0.^{\hspace*{-0.1cm}\prime\prime}35\times0.^{\hspace*{-0.1cm}\prime\prime}34$
         beam.}
         \end{figure*}
A prominent counterpart to group
(ii) is the Seyfert galaxy NGC 5643 (Fig.\,\ref{mrk266}). This
objects has similar structures as, for example, PG0953+414 (Fig.\,\ref{pg0953}).
The much higher diversity
in structure seen for NGC 5643 can be attributed to 
resolution effects. Calculating the linear scales of the radio
emission, the structures of PG0953+414 turns out to be even
larger ($\sim$5\,kpc) than those in NGC 5643
($\sim$4\,kpc; Tab.\,\ref{sample_results}). PG0026+129 is
thereby exceptional in this group as the overall structure is
$\sim$20\,kpc large.\\\indent
In group (iii), PG0157+001 (Fig.\,\ref{pg0157}) can be
compared with, for example, 
Mrk 573. Both objects consist of multiple aligned
(individually barely resolved) emitting knots in the radio regime. These radio
knots are generally interpreted as the termination points of a jet that
is stopped in the dense ISM where the (radio) emission is enhanced by
compressed material and magnetic  fields (e.g. Falcke et
al. \cite{falcke98}). The optical emission--line gas thereby displays
prominent bow--shock structures that are closely related to the radio
knots indicating strong interaction (Falcke et al. \cite{falcke98},
Ferruit et al. \cite{ferruit99}, Fig.\,\ref{pg0157}). This interaction
of a radio--jet that sweeps up material is confirmed 
spectroscopically for Mrk 573 (Ferruit et al.
\cite{ferruit99}) as well as for PG0157+001 (Leipski \&
Bennert \cite{leipski06}) by the presence of strong line--emitting
components displaced in velocity. However, it should be noted that
despite these similarities the linear distance of the interacting region 
from the nucleus is clearly larger in PG0157+001 ($\sim$3300\,pc)
than in Mrk 573 ($\sim$700\,pc).\\\indent
The jet of PG1119+120 (Fig.\,9) is comparable to the one in
 (Fig.\,\ref{mrk612}), both being part of group
(iv). The jets in 
these objects are pronounced only on one side of the nucleus and they
stand out by their bended shape. As suggested 
by Falcke et al. (\cite{falcke98}) for , the jet in 
PG1119+120 is possibly hitting a dense cloud in the ISM  
that redirects it. In contrast to the scenario suggested for 
PG0157+001 (and Mrk 573), 
the redirecting ISM cloud is too dense to produce a bow--shock (that should 
develop in a rather smooth distribution of surrounding material). 
Nevertheless, 
at the point where the redirection takes place, there is a hot spot or at 
least an ``enhancement'' in radio emission for both, PG1119+120 
(Fig.\,9) and  (Fig.\,\ref{mrk612}). Again, the linear
scales are clearly different as the 
redirection in  takes place at a distance of
$\sim$500\,pc from 
the nucleus while this point is $\sim$2500\,pc away from the nucleus
in PG1119+120. PG1612+261 can also be attributed to
group (iv) with a jet asymmetry to the western side, though the jet is
linear in this case.\\\indent 
Besides the mere morphological match, a
certain radio structure seems to be accompanied by a similar distribution of
the emission--line gas. For PG0157+001 and Mrk 573,
there is radio and NLR emission on either side of the nucleus (Falcke et
al. \cite{falcke98}, Fig.\,\ref{pg0157}). Even if
the NLR of NGC 5643 appears one--sided (Simpson et
al. \cite{simpson97}) these authors suggest that the other side of the
NLR is simply hidden by dust. In the framework of the structural match
suggested here, this would be confirmed by the two--sided radio structure
(Fig.\,\ref{mrk266}) as the radio emission is not affected by dust
obscuration. \\\indent
In contrast to the two--sided sources, objects with one--sided jets 
(or at least a clear asymmetry to one side) show this asymmetry also
in the NLR gas distribution, as found for PG1012+008 and
 (Fig.\,\ref{pg1012} and Falcke et al. \cite{falcke98},
respectively). \\\indent
To summarise, there is no significant difference between the 
radio structures of Seyfert galaxies and radio--quiet quasars. The same 
holds for the NLRs of these objects. However, the detection rate
  of structured radio emission and NLRs drops significantly with
  redshift, which can most likely be attributed to  
resolution effects as structures smaller than the beam size are
``averaged out''. This is also the reason for the higher diversity
seen in Seyferts. However, the linear scales in RQQs
are up to five times larger than that in Seyfert galaxies,
reflecting the more powerful central engine.
These arguments are supported by the fact that, at the distances
of the quasars, the radio structures 
of the Seyfert galaxies (in the same morphological group) would 
only be extended by less than an arcsecond. \\\indent
The structure of the radio emission can also be found in the 
emission--line gas where one-- or two--sidedness appear to dominate
in both regimes\footnote{While it should be noted that there are only
  two clear cases for direct radio jet--NLR interaction in our quasar sample,
the trends for one-- or two--sidedness seem to hold even at higher redshift.}.
Thus, from these morphological points of view the
RQQs can be considered  
as scaled--up versions of Seyfert galaxies. 
The power of the central engine and of the radio jet has not increased enough 
to lead to a different scenario in the more powerful RQQs. The
interaction of the radio ejecta with the emission--line gas suggested
from the corresponding images is confirmed by spectroscopic studies
(e.g. Ferruit et al. \cite{ferruit99}, Leipski \& Bennert
\cite{leipski06}). Therefore the radio jet might be responsible for
shaping the NLR gas distribution. However, dense material can alter
the shape of the radio and NLR emission significantly as proposed for
e.g. ESO428$-$G14 and PG1119+120. It should be
  noted that already Quillen et
al. (\cite{quillen99}) suggested that the morphology of the
line--emitting region depends on the distribution of dense ambient
media. It is likely that the overall structure is mostly defined by
the environmental conditions and the presence of a dense interstellar
medium. Thus, it
cannot be answered here if the radio jet is responsible for the
overall NLR morphology. Nevertheless the jet is strongly influencing the
structure on smaller scales and the velocity field of the
emission--line gas.

\subsection{On the size of the NLR and radio emission}
\begin{figure}[t]
         %\centering
         \hspace{-0.5cm}
	 \includegraphics[width=12cm]{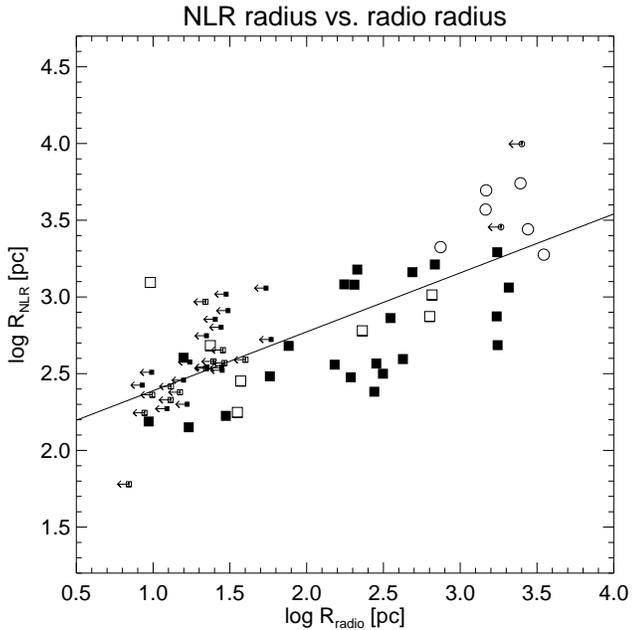}
	 \caption[]{\label{radius_radius} Correlation of the radii of
	   the [\ion{O}{iii}] and radio emitting regions.
	   Filled symbols represent type--2, open symbols
	   type--1 objects, arrows denote upper limits.
	   The RQQs data (circles) were taken from Bennert et al.
	   (\cite{bennert02}) and from this paper,
	   while the Seyfert data were taken from Falcke et
	   al. (\cite{falcke98}),  Kinney et al. (\cite{kinney00}), and
	   Schmitt et al. (\cite{schmitt01}, \cite{schmitt03}).
	   The linear
	   least square fit corresponds to $R_{\rm NLR} \propto R_{\rm
	   radio}^{0.38 \pm 0.05}$. 
	   Note that a fit excluding data points with
	   upper limits is the same within the errors.}
\end{figure}

A correlation between the [\ion{O}{iii}] luminosity and the radio 
luminosity of quasars has been known for several years (e.g. Miller et
al. \cite{miller93}). This correlation seems to hold, at least for
radio--quiet objects, if low--luminosity AGN (LLAGN) are
added (e.g. Falcke et al. \cite{falcke95}, Xu et
al. \cite{xu99}).\\\indent 
In addition to these results, the close connection of the NLR and
radio emission of Seyfert galaxies and radio--quiet quasars found here
suggests that also the sizes of these regions may be correlated among
radio--quiet objects.
Given the availability of the first direct quasar NLR sizes (Bennert et
al. \cite{bennert02}, this paper), we can now test such a relation.
We increased the database 
of our sample by additionally including the datasets
of Schmitt et al. (\cite{schmitt01}, \cite{schmitt03}) 
and Kinney et al. (\cite{kinney00}), comprising
sizes and luminosities of the [\ion{O}{iii}] 
and radio emitting regions of several Seyfert galaxies.
This sample is homogeneous
in terms of data reduction and size determination techniques which
were carried out by the same team.
The sizes and luminosities were re--calculated according
to the world model used here (\S\,3).
We also added PG1119+120 for which no direct size
  measurements of the NLR are available. However, Leipski \& Bennert
(\cite{leipski06}) have detected [\ion{O}{iii}] emission associated
with the radio knots (Fig.\,\ref{pg1119}). Thus, we estimate that the
size of the NLR in PG1119+120 is roughly the same as measured
for the radio emission.\\\indent
Combining these data, a correlation between the NLR size and radio size
is indeed   
found (Fig.\,\ref{radius_radius}), supporting the connection of both
measures. For all objects a linear least square
fit yields $R_{\rm NLR} \propto R_{\rm radio}^{0.38 \pm 0.05}$. Note
that the result of the fit does not change within the errors when
all sources with upper limits on either measure are excluded.\\\indent
Moreover, projection effects should cancel out in this correlation,
if one assumes that the axis of the ``escaping cone'' 
of ionising photons and the direction of the radio jet is the
same.
This is supported by results of Schmitt et al. (\cite{schmitt03}) who
find a very good alignment between the position angle (p.a.) of the extended
[\ion{O}{iii}] emission and that of the radio jet, confirming 
previous studies (e.g. Wilson \& Tsvetanov \cite{wilson94}).\\\indent
Given that the NLR size increases with [\ion{O}{iii}] luminosity
as implied by the NLR size--luminosity relation (Bennert et al.
  \cite{bennert02}) and assuming that the [\ion{O}{iii}] luminosity
  can be used as a measure for the total luminosity of the AGN, the
  radio emitting region becomes larger when the total luminosity is
  increased. Such a behaviour can also be seen from the combination of
  the $L_{[\ion{O}{iii}]}$ -- $L_{\rm radio}$ correlation (e.g. Miller et
al. \cite{miller93}) and the radio size--luminosity relation for
radio--quiet quasars (e.g. Morganti et al. \cite{morganti99}).  
  Thus, a possible interpretation of the observed
  correlation between NLR 
and radio size is that both depend on the intrinsic
luminosity of the central engine. However, it is not clear whether
also the interaction between radio 
jets and NLR gas influences the correlation.\\\indent
From Fig.\,\ref{radius_radius}, it is noticeable that
the radio radius increases faster than the NLR radius when
the (total) luminosity of the AGN is powered up. The radio jet is more
likely to penetrate deeper into the surrounding medium before being
stopped and disrupted (see also Barvainis et al. \cite{barvainis05}). The NLR
increases by the presence of more high--energy photons of the AGN,
that ionise the ambient gas. Thus, tuning up the central luminosity,
the size of the NLR grows while the radio jet digs its way through the
interstellar medium to overtake the NLR gas at some point.
Only for
sufficiently high jet energies the jet is likely to terminate outside
the host galaxy (Falcke et al. \cite{falckebiermann}).
\section{Conclusions}
We present deep radio images of 14 radio--quiet quasars and six Seyfert 
galaxies that show extended structures which can be interpreted as 
jets. Especially for the RQQs, aligned multiple components are detected 
that are generally interpreted as termination points of radio jets. 
The radio structures of RQQs are larger than those of Seyfert galaxies
but exhibit the same morphology.
Thus, RQQs can be regarded as scaled--up versions of Seyfert galaxies,
keeping in mind distance and resolution effects when 
interpreting the images. The similarity can also be seen in the NLR
gas distribution where the radio jet of the RQQs leaves its imprint on the NLR
creating structures  
that are well known from Seyfert NLRs. This includes sweeping--up material 
and the creation of bow-shocks (PG0157+001 vs. Mrk 573) as well as one--sided jets which interact with 
one--sided NLRs (PG1012+008 vs. ).\\\indent
Comparing our results with (radio) snapshot surveys of Seyfert galaxies and 
quasars, we have shown that these surveys can miss significant structures. 
The apparent majority of unresolved RQQs in contrast to 
the Seyferts in our dataset can be attributed to selection and
resolution effects.\\\indent
Including literature data of Seyfert NLRs and their radio emission, we
find a clear correlation between the size of the NLR and the size of the 
radio emission.
We conclude that there is no significant morphological 
difference between Seyfert galaxies 
and radio--quiet quasars on scales of typical NLR sizes (hundreds of
pc to kpc), neither in the radio nor in the [\ion{O}{iii}] emitting 
region. Moreover, the interaction of radio--ejecta with 
the NLR gas seem to be equally important in both types of sources 
for shaping the structure of the NLR {\sl and} radio emission.    

\begin{acknowledgements}
     Part of this work was supported by Sonderforschungsbereich SFB\,591
  ``Universelles Verhalten gleich-gewichtsferner Plasmen'' der
  Deutschen Forschungsgemeinschaft. We also acknowledge the comments
  of the anonymous referee.
\end{acknowledgements}

\appendix
\section{Seyfert galaxies for morphological comparison}
These six objects were selected on
         the basis of their known extended radio emission. Sources
         with various morphologies were chosen. They are bright enough to 
         be observed at 8.4 GHz to benefit from the higher angular
         resolution at this frequency.   
	 We use them to compare the structure of the {\sl extended}
         radio emission of quasars and Seyfert galaxies.
	 The higher linear resolution of the Seyfert
         maps may provide clues to the underlying structure in the
         extended quasar emission that is averaged out.
	 Any gradual transition from Seyferts to quasars
         with increasing luminosity should conserve the basic
         morphological features to some extent.
	 As already shown for the 14 RQQs and
         known from the literature,   
         the long integration times enable us to detect 
         low flux level emission of very diverse structures which is
         missed in snapshot surveys.
	 	 	      
\begin{figure*}
         \centering
         \includegraphics[width=6cm,angle=-90]{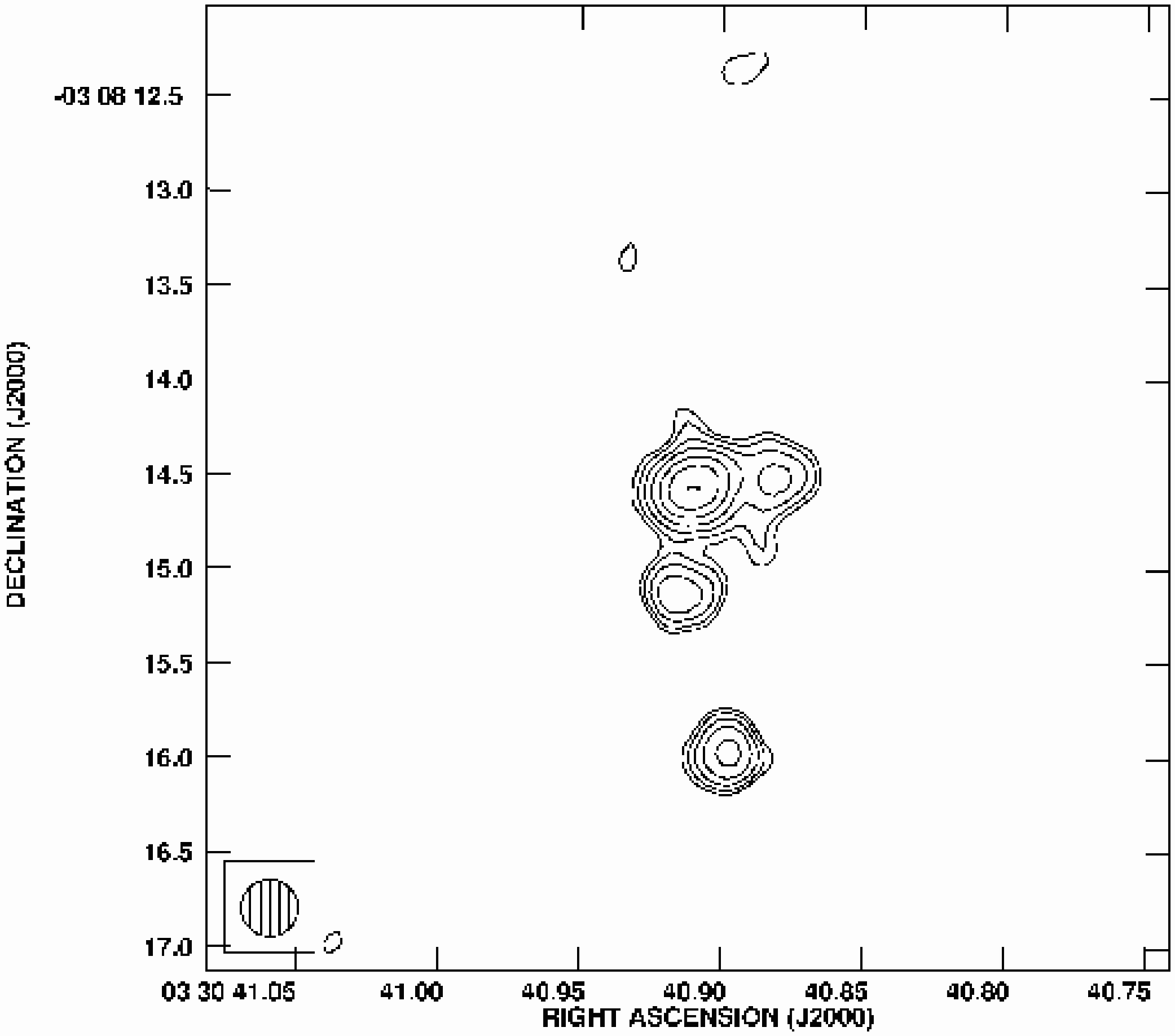}
         \includegraphics[width=6cm,angle=-90]{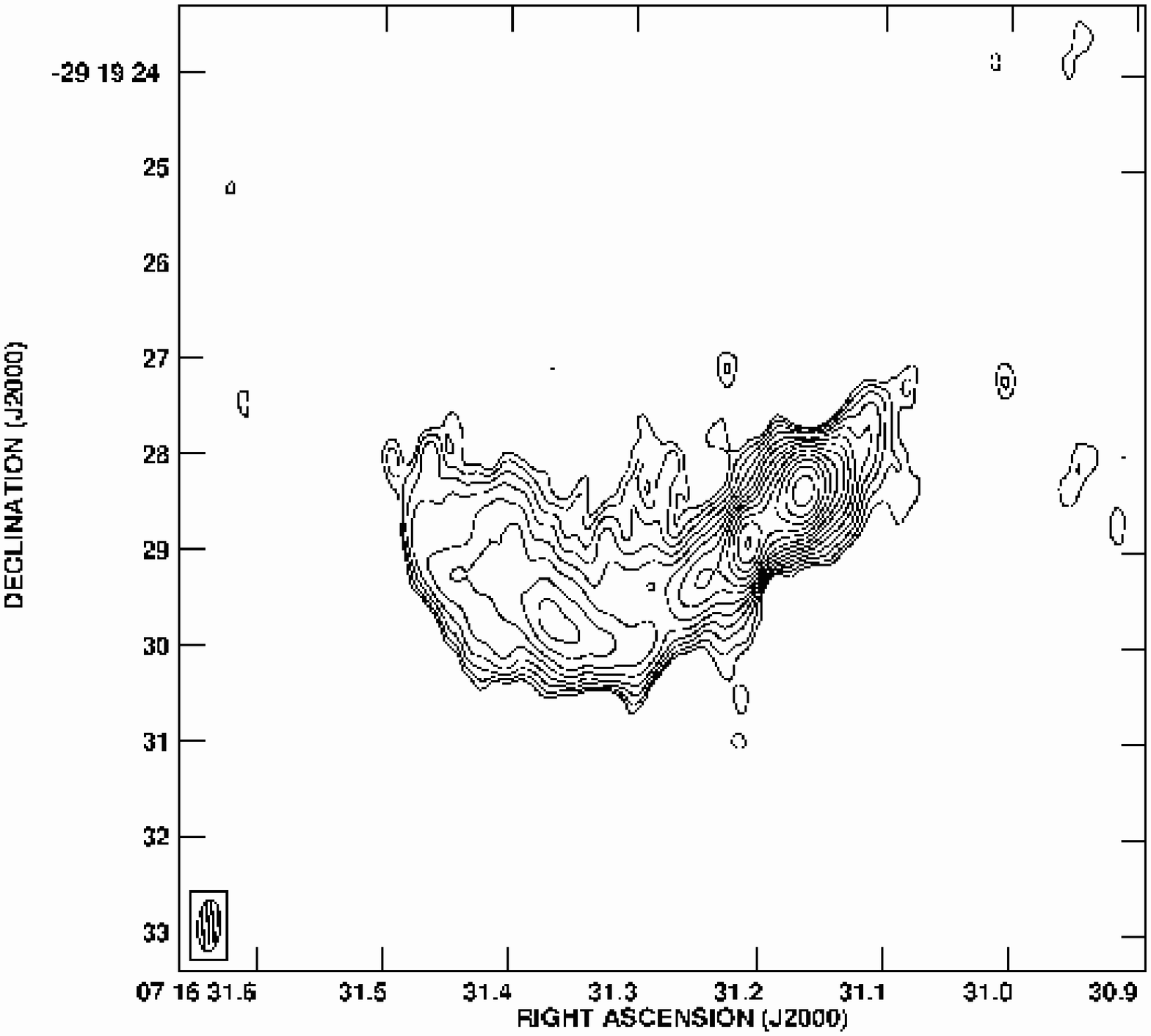}\\\hspace*{-0.5cm}
         \caption[]{\label{mrk612}VLA images of
         Mrk 612 (left) and  (right). 
         {\sl Left}: $5^{\prime\prime}\times5^{\prime\prime}$ wide
         image. Natural--weighted  
         A--Array map at 8.4 GHz with a  
         $0.^{\hspace*{-0.1cm}\prime\prime}3\times0.^{\hspace*{-0.1cm}\prime\prime}3$  
         beam.   
{\sl Right}: $10^{\prime\prime}\times10^{\prime\prime}$ wide image. 
         Natural--weighted VLA
         A--Array map at 8.4 GHz with a 
$0.^{\hspace*{-0.1cm}\prime\prime}53\times0.^{\hspace*{-0.1cm}\prime\prime}24$
         beam.}
\end{figure*}
\begin{figure*}
         \centering
         \includegraphics[width=6cm,angle=-90]{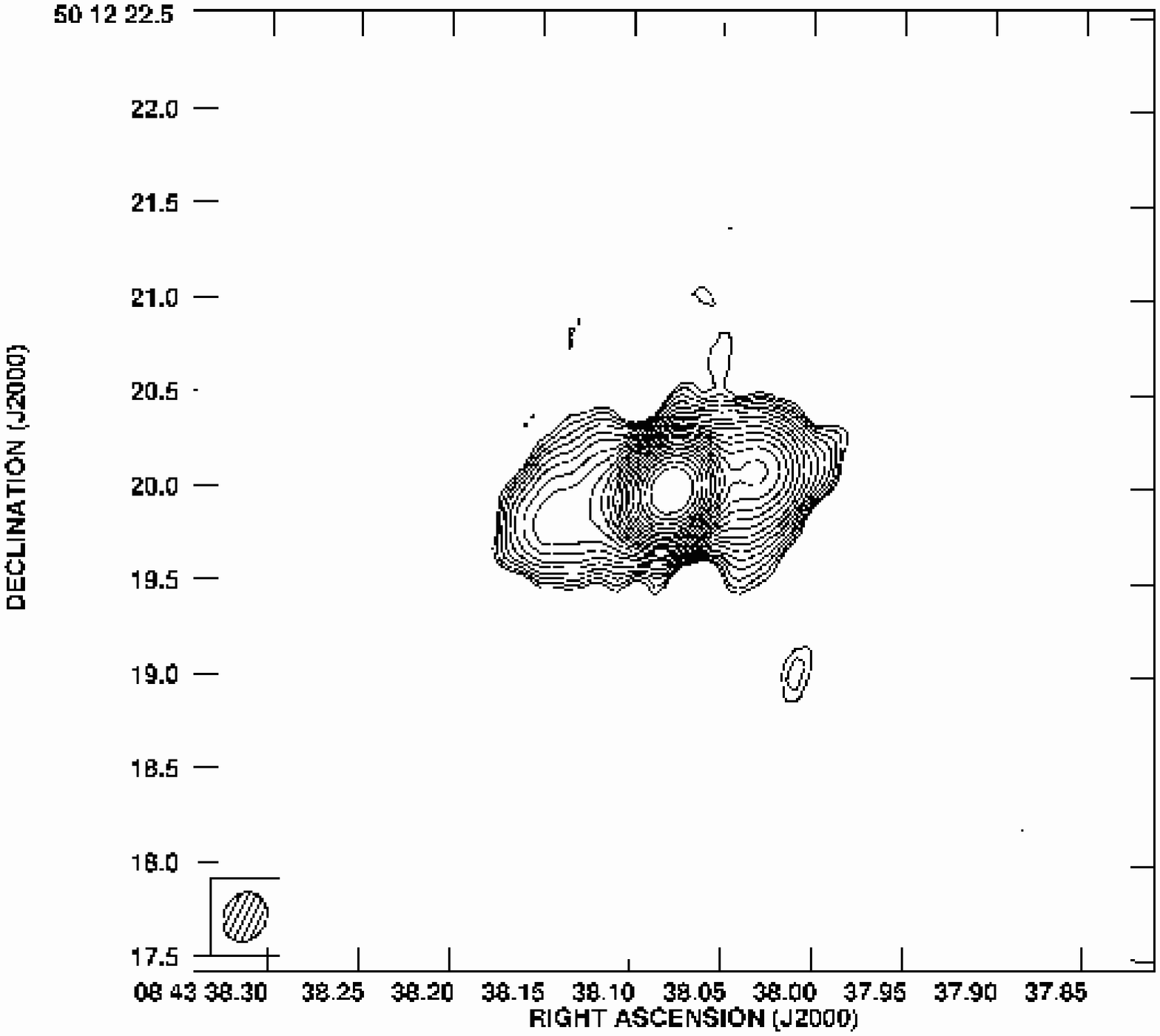}
         \includegraphics[width=6cm,angle=-90]{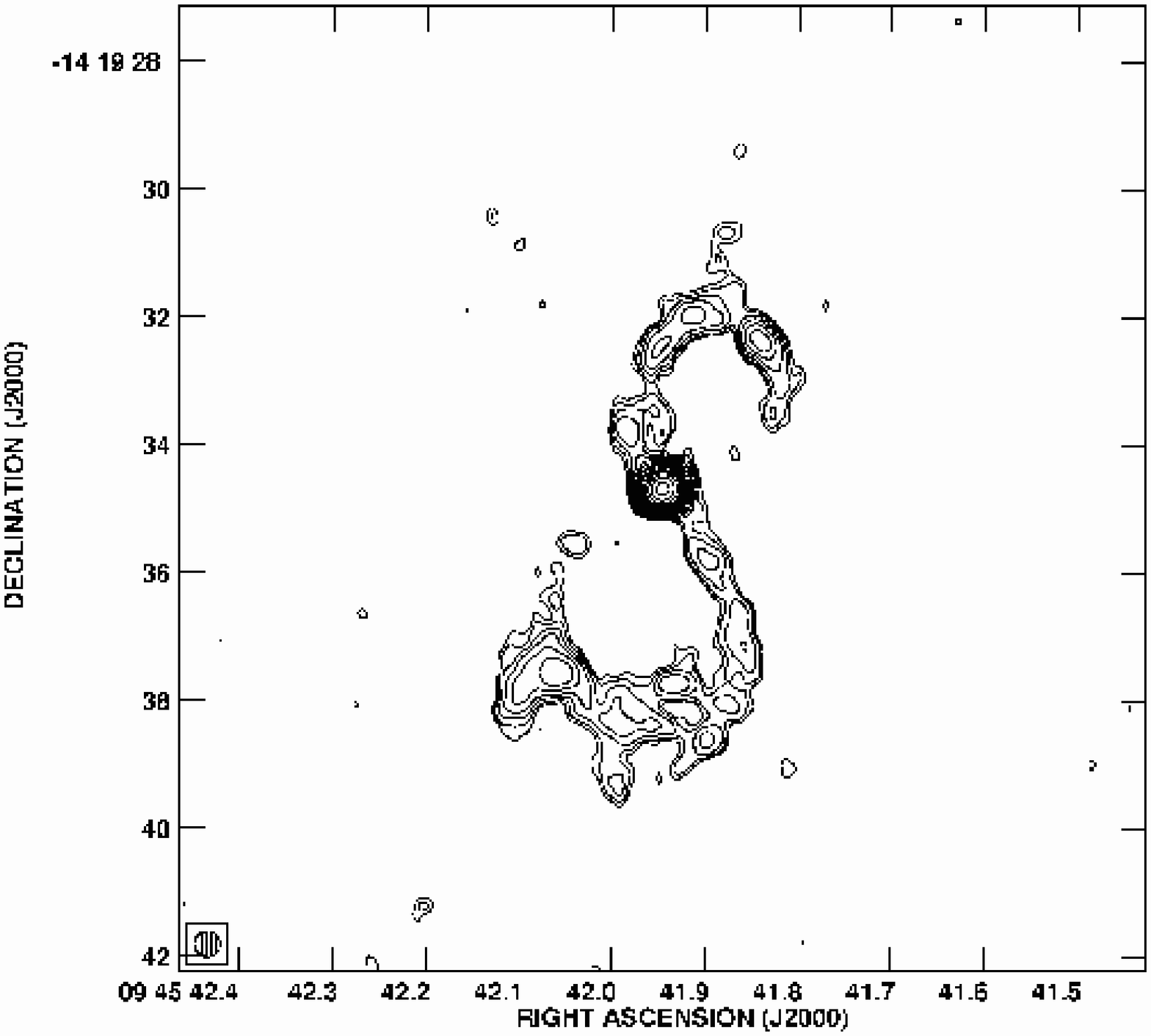}\\\hspace*{-0.5cm}
         \caption[]{\label{ngc2639}VLA images of
         NGC 2639 (left) and NGC 2992 (right). 
{\sl Left}: $5^{\prime\prime}\times5^{\prime\prime}$ wide image. Natural--weighted 
         A--Array map at 8.4 GHz with a  
         $0.^{\hspace*{-0.1cm}\prime\prime}27\times0.^{\hspace*{-0.1cm}\prime\prime}22$ 
         beam.  
{\sl Right}: $15^{\prime\prime}\times15^{\prime\prime}$ wide image. 
         Natural--weighted VLA
         A--Array map at 8.4 GHz with a 
         $0.^{\hspace*{-0.1cm}\prime\prime}4\times0.^{\hspace*{-0.1cm}\prime\prime}4$
         beam.}
\end{figure*}
{\bf Mrk 612}. Four different components can be identified in this source 
               (Fig.\,\ref{mrk612}, left). Three of them are aligned in
               north--south direction over 
               2$^{\prime\prime}$, whereas the fourth one lies closely to the 
               west of the brightest source. Nagar et
               al. (\cite{nagar99}) identified the most southern
               component to be the central one. Since the northern
               structure is elongated perpendicular to the source axis
               they interpreted this as a transverse shock.

{\bf ESO428$-$G14}. A prominent, bended, diffuse one--sided jet 
                    dominates the structure 
                    of this Seyfert--2 galaxy (Fig.\,\ref{mrk612}, right). 
                    From the central component, the jet heads towards the 
                    south--east and bends by nearly 90 degrees to the 
                    north--east after approximately 4$^{\prime\prime}$. With 
                    better resolution (but a lack of signal),
                    Falcke et al.  
                    (1998) have shown this structure also at 14.9 GHz. 
                    Our radio map and the emission--line images of Falcke 
                    et al. (\cite{falcke96a}) demonstrate the striking
                    structural similarity of  
                    radio emission and NLR emission (from HST [\ion{O}{iii}] 
                    and H$\alpha$ images) in this source.
		    
{\bf NGC 2639}. The brightest source in our sample shows a 2$^{\prime\prime}$ 
                wide structure of three components that are merged at the 
                available resolution (Fig.\,\ref{ngc2639},
                left). Interestingly, there is  
                some diffuse emission around the two outer components that is 
                constricted at the position of the central source. Wilson et 
                al. (\cite{wilson98}) reported a VLBA nucleus and the 
                variability of this source. In fact, their flux on
                arcsecond scales from
                an earlier VLA map (Ulvestad \& Wilson \cite{ulvestad89}) was 
                smaller than their VLBA flux. This flux variablity is also 
                obvious when including our data: Our peak 
                flux is 5 times greater than the flux measured by Ulvestad 
                \& Wilson (\cite{ulvestad89}) and 3 times greater than that 
                of Wilson et al. (\cite{wilson98}). Since the
                measurements in the  
                literature stated above are at 5 GHz, they are only comparable 
                with our fluxes at 8.4 GHz due to the very flat spectral 
                index of the central (dominating) source above 5 GHz (Wilson 
                et al. \cite{wilson98}).    
\begin{figure*}
         \centering
         \includegraphics[width=6cm,angle=-90]{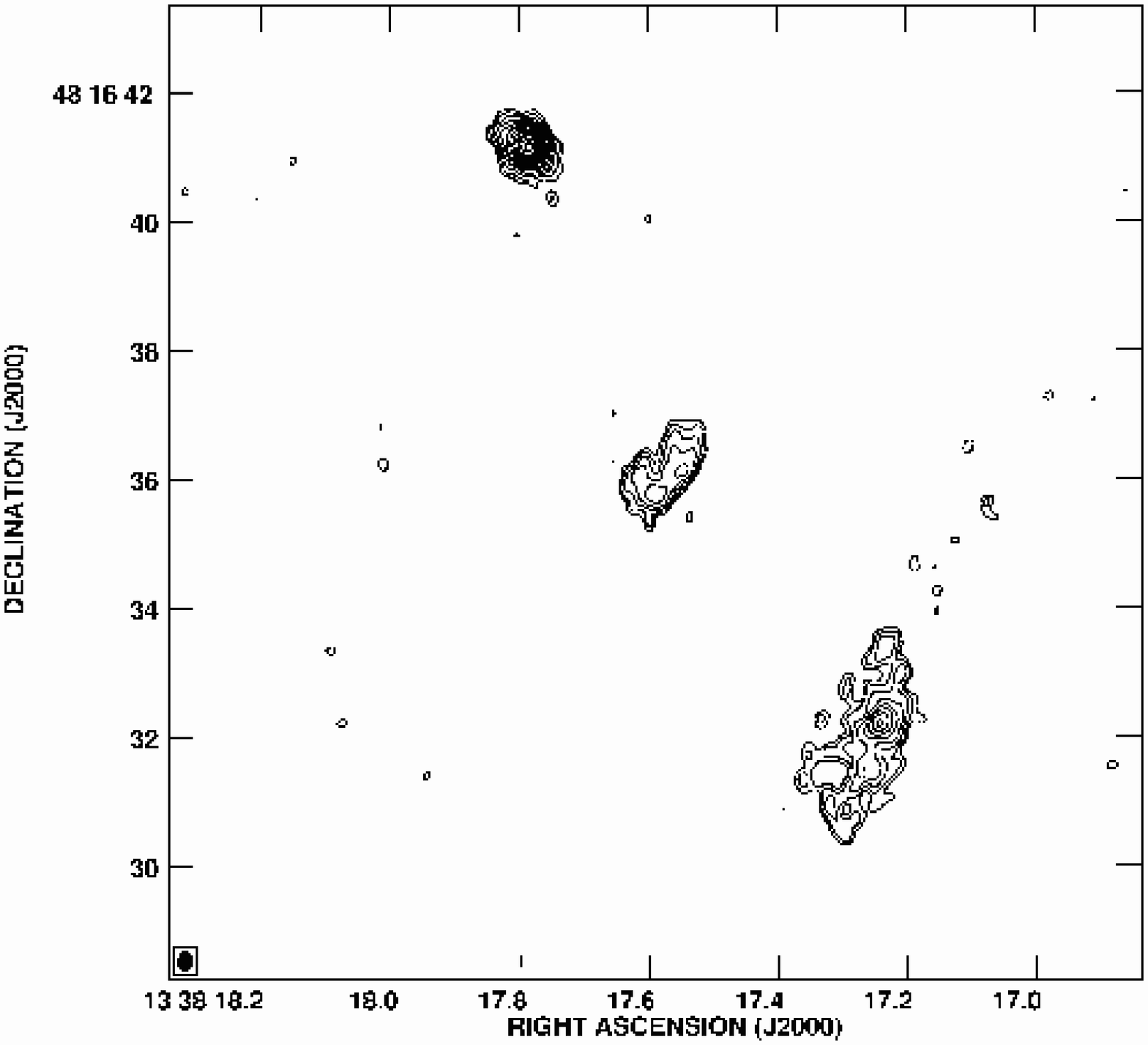}
         \includegraphics[width=6cm,angle=-90]{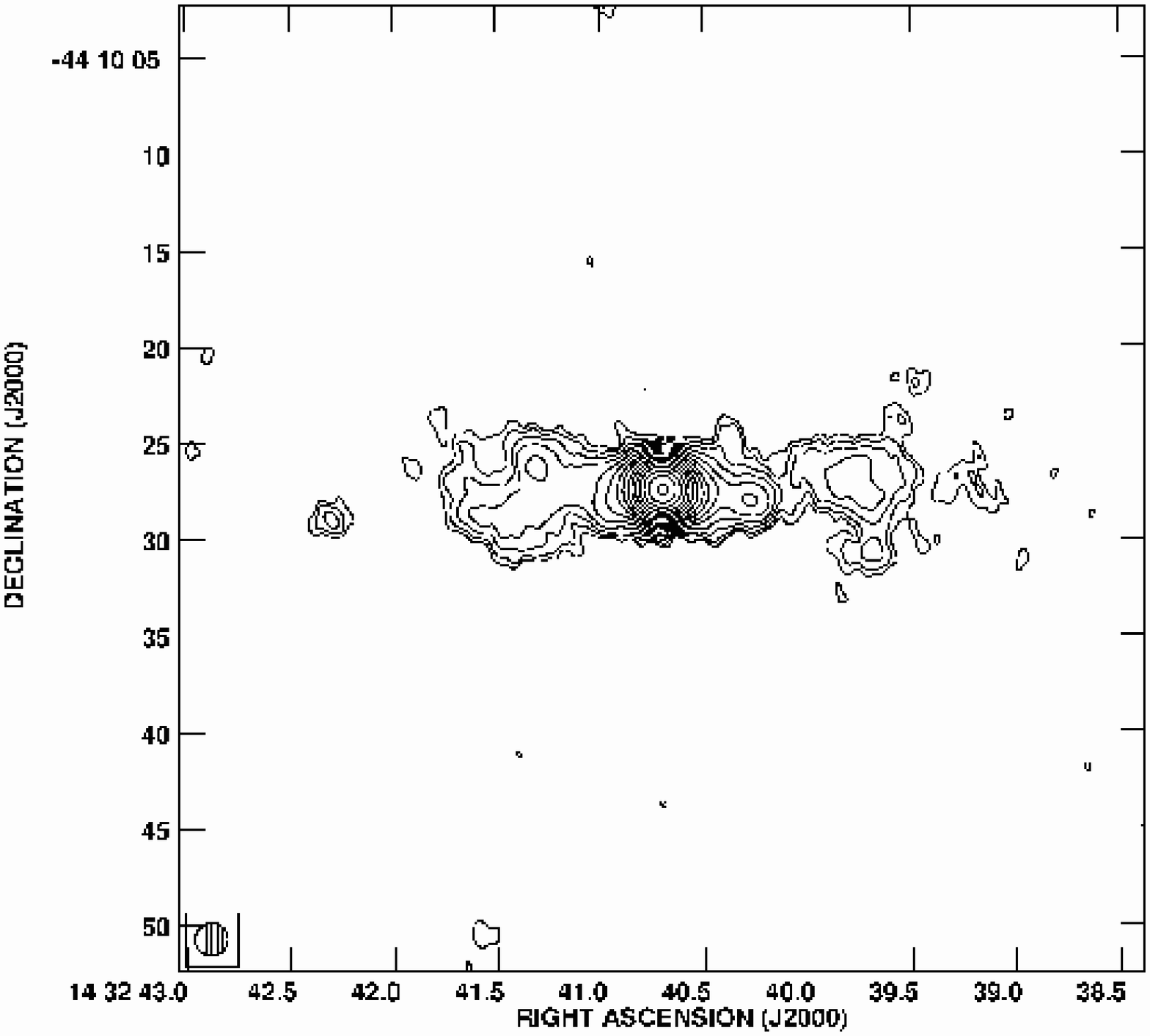}\\\hspace*{-0.5cm}
         \caption[]{\label{mrk266}VLA images of
         Mrk 266 (left) and NGC 5643 (right). 
{\sl Left}:
$15^{\prime\prime}\times15^{\prime\prime}$ wide image. Natural--weighted 
         A--Array map at 8.4 GHz with a  
         $0.^{\hspace*{-0.1cm}\prime\prime}27\times0.^{\hspace*{-0.1cm}\prime\prime}22$ 
         beam. 
{\sl Right}: $20^{\prime\prime}\times20^{\prime\prime}$ wide image.
         Natural--weighted VLA
         A--Array map at 8.4 GHz with a 
$1.^{\hspace*{-0.1cm}\prime\prime}7\times1.^{\hspace*{-0.1cm}\prime\prime}7$
         beam.}
\end{figure*}

    {\bf NGC 2992}. This object shows probably the most interesting
                structure (Fig.\,\ref{ngc2639}, right). The very prominent
                S--shaped radio emission is 
                roughly 8$^{\prime\prime}$ wide and coincides  
                with a nuclear outflow of emission--line gas (Veilleux et al. 
                \cite{veilleux01}). Their model of a hot, bipolar, thermal 
                wind that interacts with the ISM of the galaxy excludes a 
                significant contribution of the radio--jet to the energy 
                budget of the NLR. At lower (radio) frequencies (5 GHz), the 
                two loops are closed and a figure--8 structure appears 
                (Ulvestad \& Wilson \cite{ulvestad84}).

{\bf Mrk 266}. This source consists of three components: one compact
               core as well as 
               two diffuse emission regions (Fig.\,\ref{mrk266},
               left). All sources are  
               well aligned and the whole structure is roughly 
               12$^{\prime\prime}$ long. This galaxy
               with a double--nucleus in the optical shows LINER
               emission in the north--eastern nucleus 
               whereas the south--western nucleus is a Seyfert--2 type
               object (Mazzarella et al. \cite{mazzarella88}). They
               conclude that this galaxy is a merger of a
               Seyfert and a LINER galaxy. The north--eastern radio
               component most likely belongs to the LINER
               nucleus. This interpretation is strengthened by the
               fact that compact 
               radio structures are well know for LINER galaxies
               (e.g. Nagar et al. \cite{nagar00}). On the other hand,
               the south--western structure belongs to the Seyfert
               galaxy. The radio structure of this region with
               double--sided jet--like diffuse emission is 
               very typical for Seyfert galaxies. The diffuse emission
               between both components is considered
               to be the region of interaction with enhanced
               synchrotron emission.  

{\bf NGC 5643}. This source has a diffuse radio jet on both sides
                of the nucleus which is nearly 
                30$^{\prime\prime}$ long (Fig.\,\ref{mrk266},
                right). The whole  
                radio structure matches well with the corresponding 
                distribution of the NLR gas (Simpson et al. 
                \cite{simpson97}). Although Simpson et al. 
                (\cite{simpson97}) detect only a
                one--sided ionisation cone on the eastern side of the
		nucleus they state that the western side 
                suffers heavy extinction by a dust lane, thus
		shadowing the second cone.

\end{document}